\documentclass[11pt,draftcls,onecolumn]{IEEEtran}
\usepackage{amsfonts, amsmath, amssymb}
\usepackage[buvect]{commands09}
\usepackage{vect09,matrix09}
\usepackage{cite}
\usepackage{color}
\usepackage{subfigure}
\usepackage{graphicx}
\usepackage{algorithm}
\usepackage{algorithmic}
\usepackage{verbatim}
\usepackage{paralist}

\newtheorem{proposition}{Proposition}

\newcommand{\mmt}{M}            % # of measurements
\newcommand{\mmi}{m}            % index
\newcommand{\len}{N}            % # of unknowns
\newcommand{\lei}{n}            % index
\newcommand{\nis}{\W}           % the noise vector
\newcommand{\thr}{T}

\newcommand{\snc}{\gamma}

\IEEEoverridecommandlockouts

\newcommand{\LZ}[1]{{\color{black} #1}}
\newcommand{\LZnew}[1]{{\color{black} #1}}
\newcommand{\DG}[1]{{\color{black} #1}}

\newcommand{\ZH}[1]{{\color{black} #1}}
\newcommand{\ind}[1]{1\hspace{-2mm}{1}\left(#1\right)}
\newcommand{\diff}{{\rm d}}

\begin{document}

\title{Neighbor Discovery for Wireless Networks via Compressed Sensing}

\author{Lei Zhang, Jun Luo and Dongning Guo}

\maketitle

\begin{abstract}
  This paper studies the problem of neighbor discovery in wireless
  networks, namely, each node wishes to discover and identify the
  network interface addresses (NIAs) of those nodes within a single
  hop.  A novel paradigm, called {\em compressed neighbor discovery}
  is proposed, which enables all nodes to simultaneously discover
  their respective neighborhoods with {\em a single frame} of
  transmission, which is typically of a few thousand symbol
  epochs. The key technique is to assign each node a unique {\em
    on-off signature} and let all nodes simultaneously transmit their
  signatures.  Despite that the radios are half-duplex, each node
  observes a superposition of its neighbors' signatures (partially)
  through its own off-slots.  To identify its
  neighbors out of a large network address space, each node solves a
  {\em compressed sensing} (or {\em sparse recovery}) problem.

  Two practical schemes are studied.  The first employs
  random on-off signatures, and each node discovers its
  neighbors using a noncoherent detection algorithm
  based on group testing.
  The second scheme uses on-off signatures based on a deterministic
  second-order Reed-Muller code, and applies a chirp decoding
  algorithm.  The second scheme needs much lower
  signal-to-noise ratio (SNR) to achieve the same error performance.
  % A network of over one million Poisson distributed nodes (with 20-bit
  % NIAs) is studied numerically, where each node has 30 neighbors on average, and
  % the channel between each pair of nodes is subject to path loss and
  % Rayleigh fading.  Within a single frame of 4,096 symbols, nodes
  % can discover their respective neighbors with on average 99.7\%
  % accuracy at 10 dB SNR.
  The complexity of the chirp decoding algorithm is
  sub-linear, so that it is in principle scalable to  networks with billions of
  nodes with 48-bit IEEE 802.11 MAC addresses.
  The compressed neighbor discovery schemes are much more efficient
  than conventional random-access discovery, where nodes have to retransmit
  over many frames with random delays to be successfully
  discovered.
\end{abstract}


\section{Introduction}\label{sec:ND}

In many wireless networks, each node has direct radio link to only a
small number of other nodes, called its {\em neighbors (or peers)}.
Before efficient routing or other network-level activities are
possible, nodes have to discover and identify the network interface
addresses (NIAs) of their neighbors.  This is called {\em neighbor
  discovery (or peer discovery)}.  The problem is crucial in mobile ad
hoc networks (MANETs), which are self-organizing networks without
pre-existing infrastructure.  The problem is becoming
important in increasingly more heterogeneous cellular networks with
the deployment of unsupervised picocells and femtocells.

A node interested in its neighborhood, which is henceforth referred to as the {\em query node}, listens to the wireless channel during the discovery period, and then decodes the NIAs of its neighbors. Neighbors transmit signals which contain their identity information. It is fair to assume that non-neighbors either do not transmit, or their signals are weak enough to be regarded as noise.  We make two important observations: 1) The physical channel is a multiaccess channel, where the observation made by the query node is a (linear) superposition of transmissions from its neighbors,
corrupted by noise; 2)
% Regardless of the signalling format,
The goal of neighbor discovery is to identify, out of all valid NIAs,
which ones are used by its neighbors.

%{\color{blue} In such a scenario, it is crucial to deal with the near-far problem where a strong signal captures a receiver making it impossible for the receiver to detect a weaker signal.}

%  The received
%signal, often slotted, can be regarded as a certain number of
%measurements over time, one measurement per slot.

State-of-the-art neighbor discovery protocols, such as that of the
IETF MANET working group~\cite{RFC-TBRPF} and the ad hoc mode of IEEE
802.11 standards, can be described as follows: The query node
broadcasts a probe request.  Its neighbors then reply with probe
response frames containing their respective NIAs.  If a response frame
does not collide with any other frame, the corresponding NIA is
correctly received.  Due to lack of coordination, each
neighbor has to retransmit its NIA enough times with random delays, so
that it can be successfully received by the query node with high
probability despite collisions. We refer to such a scheme as {\it
  random-access neighbor discovery}.
Several such algorithms which operate in or on top of medium access
control (MAC) layer have been proposed~\cite{McGBor01MANC,
  VasKurInfocom05, FelEki10SECON, VasTow09ICMCN, KhaGoe10Infocom,
  BorEph07AdhocNet}.
%, CorLyn09Allerton, KohSuh09AdhocNet}.

\begin{comment}
There are several such algorithms operated in or on top
of medium access control (MAC) layer proposed in the literature such
as \LZ{the} ``birthday protocol"~\cite{McGBor01MANC}, directional antenna
neighbor discovery~\cite{VasKurInfocom05,FelEki10SECON},
among others~\cite{CorLyn09Allerton, KohSuh09AdhocNet}.
%neighbor discovery using an abstract MAC layer~\cite{CorLyn09Allerton} and
%energy-efficient neighbor discovery for low-duty cycle MAC
%schemes~\cite{KohSuh09AdhocNet}, etc.
\end{comment}

%It is easy to see that all nodes can in fact {\em simultaneously} discover their respective neighborhoods and transmit their NIAs during the discovery period.

Random access assumes a specific
signalling format, namely, a node's response over the discovery period
basically consists of repetitions of its NIA interleaved with periods
of silence. This signalling format allows the NIA to be directly read
out from a successfully received frame.
\DG{Every node can discover its neighborhood and also be
  discovered by neighbors given long-enough discovery period.}
However, such signalling is far from optimal.  To design the optimal
signalling, we should remove all unnecessary structural
restrictions on the responses. Given the duration of the discovery
period, the problem is in general
to assign each node a distinct response, or {\em signature} over that period, and
to design a decoding algorithm for a query node
to identify the constituent signatures (or corresponding NIAs) based
on the observed superposition. It would be ideal if all the signatures
were orthogonal to each other, but this is impossible in case
the number of signatures far exceeds the signature length.  A good
design should
make the correlation between any pair of signatures as small as possible.

A crucial observation is that the number of actual neighbors is typically
orders of magnitude smaller than the node population, or more
precisely, the size of the NIA space, so that neighbor discovery is by
nature a {\em compressed sensing (or sparse recovery)}
problem~\cite{Donoho06IT, CanTao06IT}.  By the wisdom from the compressed
sensing literature, the required number of measurements (the signature
length) is dramatically smaller than the size of the NIA space.

Based on the preceding observations, this work provides a novel
solution, referred to as {\em compressed neighbor discovery}, which
attains highly desirable trade-off between reliability and the length
of the discovery period, thus minimizing the neighbor discovery
overhead in wireless networks.
The defining feature is to let nodes simultaneously transmit their
signatures within a single frame interval. \LZ{In order to let each
  node discover its own neighborhood during the same frame interval it
  is transmitting, i.e., to achieve full-duplex neighbor discovery,
  the signatures consist of on- and off-slots, so that within the
  discovery frame a node can make observations during its off-slots
  and also transmit during its on-slots.}  Some sparse recovery
algorithm is then carried out to decode the neighborhood.

\DG{The organization of the remaining sections of the paper and our
  key contributions are as follows.}
After the system model is presented in Section II, two types of
signatures with corresponding decoding
algorithms are proposed. The first scheme, which is studied in
Section~\ref{s:onoff}, assigns each node a pseudo-random {\em on-off
  signature} (i.e., a sequence of delayed pulses) over the (slotted)
discovery frame. The number of on-slots is a small fraction of the
total number of slots, so that the signature is sparse.  The
superposition of the signatures of all neighbors is a denser sequence
of pulses, in which a pulse is seen at a slot if at least one of the
neighbors sent a pulse during the slot. A simple decoding
procedure via eliminating non-neighbors is developed based on
algorithms originally introduced for {\em group
  testing}~\cite{DuHwa93, BerMeh84TC}. The complexity of the algorithm
is linear in the address space, which is feasible for networks with
moderately large but not too large NIA spaces.

The second scheme, which is studied in Section~\ref{s:rm}, generates a
set of deterministic signatures based on a second-order Reed-Muller
(RM) code.  First- and second-order RM codes date back to 1950s and
%~\cite{ZhaGuo11Wiopt}.
are fundamental in the study of error-control codes and
algorithms~\cite{Sudan01FOCS}.
More recently, RM codes have been shown to be excellent
for sparse recovery~\cite{HowCal08CISS}. The original RM code consists
of quadrature phase-shift keying (QPSK) symbols, with no
  off-slots.  In order to achieve full-duplex discovery, we
introduce off-slots by replacing roughly a half of the QPSK symbols by zeros.
The chirp decoding algorithm of~\cite{HowCal08CISS} is modified to
perform despite the erasures.
The choice of modified RM codes for neighbor discovery is not
incidental: The algebraic structure allows unusually low decoding
complexity (sublinear in the number of codewords), so that the
scheme is in principle scalable to $2^{48}$ or more %$2^{65}$
nodes or NIAs in the network~\cite{ZhaGuo11Wiopt}.

In Section~\ref{s:Compare}, compressed neighbor discovery is compared
with random-access schemes and shown to require much fewer
transmissions to achieve the same error performance. In addition, the new scheme entails much less transmission overhead (such as preambles and parity checks), because it takes a single frame of transmission, as opposed to many frame transmissions in random access.

We highlight some of the unique contributions of this work:
\begin{compactitem}
\item It is the first to propose on-off signalling for achieving
{\em full-duplex} neighbor discovery using half-duplex radios,
  which departs from conventional schemes where the transmitting
  frames of a node are scheduled away from its own receiving frames;
\item This work is the first to use Reed-Muller codes for neighbor
 discovery, which enables highly efficient discovery for networks of any practical size;
\item Previous work~\cite{LuoGuo08Allerton, LuoGuo09Allerton} only
  models the neighborhood of a single query node.  This paper
  considers a more realistic network modeled by a Poisson point
  process, and a more realistic propagation model with path loss;
\item The decoding algorithm for random on-off signatures
significantly improves the performance
  of the group-testing-based algorithms studied
  in~\cite{LuoGuo08Allerton} and~\cite{LuoGuo09Allerton} for noiseless
  and Rayleigh fading channels, respectively;   %Bounds are also
  %developed for the error probability in order to optimize the design;
\item Previous work~\cite{LuoGuo09Allerton} only
  demonstrates reliable discovery at high signal-to-noise ratio (SNR)
  for a rather sparse network in which the average number of neighbors
  is less than ten.  Numerical results in this paper demonstrate
  reliable and efficient discovery of 30 neighbors or more at fairly
  low SNR.

% \item Important design issues such as synchronized transmission of the
%   signatures and the near-far effect are also addressed in the remainder of this paper.
\end{compactitem}

%%%%%%%%%%%%%%%%%%%%%%%%%%%%%%%%%
\section{The Channel and Network Models}\label{s:model}

\subsection{The Linear Channel} % (or Measurement System)}

Consider a wireless network where each node is assigned a unique
network interface address.  Let the address space be
$\{0,1,\dots,N\}$ (e.g., $N=2^{48}-1$ if the space consists of all
IEEE 802.11 MAC addresses).
The actual number of nodes present in the network
can be much smaller
than $N$, but as far as neighbor discovery is concerned,
we shall assume that there are exactly $N+1$ nodes.

We will later discuss the problem of having all nodes simultaneously
discover their respective neighborhoods, but for now let us assume
that node $0$ is the only query node and sends a probe signal to
prompt a neighbor discovery period of $M$ symbol intervals.
Each node $\lei$ in the
neighborhood responds by sending a signal $\S_\lei = [S_{1\lei},\dots,S_{\mmt\lei}\tran{]}$.
The signal identifies node $\lei$ and
is also referred to as the {\em signature} of node $\lei$.
In case a node only transmits over selected time instances,
those symbols $S_{mn}$ corresponding to non-transmissions are
regarded as zero.
For the time being let us ignore the variation of the small
propagation delays between the query node and its neighbors, and
assume symbol-synchronous transmissions from all nodes.
We also assume that this discovery period is shorter than the channel coherence time.
The received signal of node $0$ can thus be expressed as
\begin{equation}\label{eq:Y}
  \Y = \sqrt{\snc} \sum_{\lei\in \N_0} U_\lei \S_\lei + \nis
\end{equation}
where $\N_0$ denotes the set of NIAs in the neighborhood of node
$0$, $U_\lei$ denotes the complex-valued coefficient of the
wireless link from node $n$ to node 0, $\gamma$ denotes the average
channel gain in the SNR, and $\nis$ consists
of $\mmt$ independent unit circularly symmetric complex Gaussian
random variables, with each entry $W_\mmi \sim \mathcal{CN}(0,1)$. For
simplicity, transmissions from non-neighbors, if any, are
accounted for as part of the additive Gaussian noise.

% It is fair to assume that the signature\DG{s} of \DG{all nodes are}
% % each node is
% pre-determined.
% %and unrelated to any other nodes.
% Suppose also that all signatures are known to node $0$.
The goal
 is %thus
to recover the set $\N_0$,
given the observation $\Y$, the SNR $\gamma$,
and knowledge of the signatures $\S_1,\dots,\S_\len$.
The random coefficients $U_\lei$ are unknown except for its statistics.
For convenience, we introduce binary variables $B_\lei$,
which is set to $1$ if node $\lei$ is a neighbor of node 0, and set to $0$ otherwise.
% The neighborhood set $N_0$ is then equivalent to a
% binary vector of length $\len$, denoted as $\B=[B_1,\dots,B_\len\tran{]}$.
Let $\X=[B_1U_1, \dots, B_\len U_\len\tran{]}$ and
$\SSS = [\S_1, \dots, \S_\len]$.
Then model~\eqref{eq:Y} can be rewritten as
\begin{equation} \label{eq:YS}
    \Y = \sqrt{\snc} \SSS \X + \nis
\end{equation}
where we wish to determine which entries of $\X$ are nonzero, i.e., to
recover the support of $\X$. %, for given $\Y$ and $\SSS$.

Model~\eqref{eq:YS} represents a familiar noisy linear measurement
system.  We shall refer to $\Y=[Y_1,\dots,Y_\mmt\tran{]}$ as the
measurements, and $\SSS_{\mmt\times\len}$ as the known signature matrix.
It is reasonable to assume that $B_1,\dots,B_\len$ are independent and
identically distributed (i.i.d.) Bernoulli random variables with
$\Prob\{B_1=1\}=c/\len$, where $c$ denotes the average number of
neighbors of node $0$.
% WE DO NOT LIMIT TO ON-OFF SIGNALLING HERE
% \JL{ In addition, we generate the measurement matrix $\SSS$ such that
%   $S_{\mmi,\lei}$ are independent Bernoulli($q$) random variables
%   where the parameter $q$ is to be designed.}
Let us further assume that $U_1,\dots,U_\len$ are i.i.d.\ with known
distribution, % $P_U$,
and are independent of $B_1,\dots,B_N$ and noise.  To recover the support of
$\X$ is then a well-defined, familiar statistical inference problem.

The node population $\len+1$ is typically much larger than the number
of symbol epochs in one discovery period $\mmt$, so that the
  linear system~\eqref{eq:YS} is under-determined even in the absence
  of noise.  An important observation
is that the vector variable $\X$ is very sparse,
so that neighbor  discovery is fundamentally a sparse recovery
problem, which implies that
very few measurements, which can be orders of magnitude smaller than
$\len$, are sufficient for reconstructing the $N$-vector $\X$ or its support~\cite{WuVer10IT}.

\subsection{Signatures and Their Distribution}

In the case of random-access neighbor discovery, each $\S_\lei$
consists of repetitions of the NIA of node $\lei$ interleaved with
random delays, sufficient synchronization flags, training symbols and
parity check bits are embedded so that the delays can be measured
accurately (this constitutes substantial overhead).

In general, the signature of node $n$ can be regarded as the $n$-th
codeword from the codebook $\SSS$.  (In case of delay uncertainty, we can
use a larger codebook to include shifted versions of the signatures.)
The signatures of all nodes, i.e., the codebook, should be known input
to the neighbor discovery algorithm carried out by any query node.

To make the distribution of a large codebook to all nodes practical, some simple
structure shall always be introduced.  For example, the signature of node
$n$ can be generated using a common pseudo-random number generator
with the seed equal to $n$.  It then suffices to distribute the generator
(e.g., as a built-in software/hardware function) in lieu of the
signatures.  In principle, each node can construct the codebook $\SSS$ by
enumerating all valid NIAs, so that all signatures are known to all
nodes in advance without any communication overhead.  It is also
possible to design an inverse
mapping to recover the index $n$ given any signature $\S_n$.
An alternative design is to let the signatures be codewords of an
error-control code, in which case it suffices to reveal the code to all nodes.

A key finding of this paper is that, in order for efficient neighbor
discovery, the signatures $\S_\lei$ should not merely consist of repetitions
of the NIA.  %That is, random-access discovery is not efficient.
Discovery using cleverly designed signatures is not only feasible, but
can be significantly more efficient than random-access discovery.

% In practice, the signature of each node can be generated using a
% common pseudo-random number generator with its own NIA as the seed,

\subsection{Propagation Delay and Synchronicity} %Model} % and Neighborhood }

In general, a receiver has to resolve the timing uncertainty of its
neighbors in order to recover their identities.
% At any rate, there is no fundamental disadvantage by using compressed
% neighbor discovery, because any alternate scheme has to deal with the
% same timing uncertainty.
By including sufficient synchronization flags, random-access schemes are robust with respect to random delays.
Since it is costly to add enough redundancy to allow accurate
estimation of the delays in a multiuser environment, it can be
beneficial to let nodes transmit their signatures simultaneously and
synchronously.
Some common clock, such as access to the
global positioning system (GPS) can provide the timing needed. In our scheme,
it suffices to have all communicating peers be approximately
symbol-synchronized, as long as the timing difference (including the
propagation delay) is much smaller than the symbol interval. This can
be achieved by using distributed algorithms for reaching average
consensus~\cite{SimSpa08SPM}.

By definition neighbors should be physically close to the query node, so that
the radio propagation delay is much
smaller compared to a symbol epoch.  For instance, if neighbors are
within $300$ meters, the propagation delay is at most $1$ microsecond, which
is much smaller than the bit or pulse interval of a typical MANET.
More pronounced propagation delays can also be explicitly addressed in
the physical model, but this is out of the scope of this paper.

Admittedly, synchronizing nodes requires an upfront cost in the
operation of a wireless network.  The benefit, however, is not limited
to the ease of neighbor discovery, but improved efficiency in many
other network functions.  Whether synchronizing the nodes is
worthwhile is a challenging question, which is
not discussed further in this paper.

\subsection{Propagation Loss and Near-Far Problem}
In previous work~\cite{LuoGuo09Allerton}, we considered a single query node and neighbors of the same distance, and simply assumed the channel gains $U_\lei$ to be Rayleigh fading random variables. In this paper, we incorporate the effect of network topology and propagation loss in the channel model. Suppose all nodes transmit at the same power, large-scale attenuation follows power law with path loss exponent $\alpha$, and small-scale attenuation follows i.i.d.~fading.  Due to reciprocity, the gains of the two directional links between any pair of nodes are identical.

From the viewpoint of a query node, it suffices to describe the statistics of
$U_\lei$ \ZH{of neighboring nodes} in model~\eqref{eq:Y} as follows.
%Let $N$ nodes be (``uniformly'') distributed on a large area
%of size $A$ according to a homogeneous Poisson point process.  We
%ignore the boundary effect.
Suppose all nodes are distributed in a plane according to a homogeneous Poisson point process with intensity $\lambda$. Consider a uniformly and randomly selected pair of nodes. The channel power gain between them is $GR^{-\alpha}$, where $G$ denotes small-scale fading and $r$ stands for the distance between them.  The nodes are called neighbors of each other if the channel gain between them exceeds a certain threshold, i.e., $G R^{-\alpha}>\eta$ for some fixed threshold $\eta$. We choose not to define the neighborhood purely based on the
geometrical closeness because: 1) connectivity between a pair of nodes is determined by the channel gain; and 2) a receiver cannot separate the attenuations due to path loss and Rayleigh fading in one discovery period.

\ZH{Consider an arbitrary neighbor, $n$, of the query node, where the distance between them is $R$, and the random attenuation of the channel is $G$. By definition of a neighbor, $G$ and $R$ must satisfy $GR^{-\alpha}\geq\eta$, i.e., $R\leq\left(G/\eta\right)^{1/\alpha}$. Under the assumption that all nodes form a Poisson point process, for given $G$, this arbitrary neighbor $n$ is uniformly distributed in a disc centered at the query node with radius $\left(G/\eta\right)^{1/\alpha}$. Therefore, the conditional distribution of $R$ given $G$ can be expressed as
\begin{align} \label{eq:condProb}
\Prob(R\leq r\big|G) = \left\{
    \begin{array}{ll}
      r^2\left(\frac{\eta}{G}\right)^\frac{2}{\alpha}, & r\leq \left(\frac{G}{\eta}\right)^\frac{1}{\alpha}; \\
      0, & \text{otherwise}.
    \end{array}
    \right.
\end{align}
Now for every $u \geq \sqrt{\eta}$, by~\eqref{eq:condProb} we have
\begin{align}
\Prob(GR^{-\alpha}\geq u^2) &= \expsub{G}{\Prob\left(R \leq \left(\frac{G}{u^2}\right)^{\frac{1}{\alpha}} \bigg|G \right)} \nonumber \\
&= \expsub{G}{\left(\frac{\eta}{u^2}\right)^\frac{2}{\alpha}} \nonumber \\
&= \frac{\eta^{\frac2\alpha}}{u^{\frac4\alpha}}\,.
\end{align}
}
\begin{comment}
The distribution of the amplitude $|U_\lei|$ \ZH{of any neighboring node}
is identical to that of $\sqrt{Gr^{-\alpha}}$ conditioned on that it exceeds $\sqrt{\eta}$.
For every $u \geq \sqrt{\eta}$,
\DG{
\begin{align}
  \label{eq:6}
  \Prob(|U_n| > u \big| Gr^{-\alpha} > \eta) =
  \frac{ \Prob(Gr^{-\alpha} > u^2 )}{ \Prob(Gr^{-\alpha} > \eta) } =
 \frac{u^{-\frac4\alpha}}{\eta^{-\frac2\alpha}}
\end{align}
}
\DG{because
  \begin{align}
  \Prob( G r^{-\alpha} > \eta )
   &= \expect{ \Prob\left( r
       <\left(\frac{\eta}G\right)^{-\frac1{\alpha}} \Big|G \right)
   } \label{eq:Gr} \\
   &= \expect{ \frac{\pi}A \left( \frac{\eta}G \right)^{-\frac2{\alpha}}}
\end{align}
where the probability in~\eqref{eq:Gr} is evaluated as the probability
that an arbitrary node is found on a disk centered at the query node.
}
\end{comment}
Hence the probability density function (pdf) of $|U_\lei|$ of neighbor $n$ is
\begin{align} \label{eq:Un_pdf}
  p(u) = \left\{
    \begin{array}{ll}
      \frac{4}{\alpha} \frac{\eta^{2/\alpha}}{u^{4/\alpha+1}}, & u \geq \sqrt{\eta}; \\
      0, & \text{otherwise}.
    \end{array}
    \right.
\end{align}
Interestingly, the distribution does not depend on the
fading statistics (of $G$).  Moreover,
it is fair to assume that the coefficients
are circularly symmetric, i.e., the phase of $U_n$ is uniform on
  $[0,2\pi)$.

Without loss of generality, we assume that the query node locates at the origin. Denote $\Phi=\{X_i\}_i$ as the point process consisting of all nodes excluding the query node. By Slivnyak-Meche theorem~\cite{BacBla09FT}, $\Phi$ is a Poisson point process with intensity $\lambda$. Fading coefficients can be regarded as independent marks of the point process, so that $\tilde{\Phi}=\{(X_i,G_i)\}_i$ is an independently marked Poisson point process. By Campbell's theorem~\cite{BacBla09FT}, the average number of the query node can be obtained as:
\begin{align}
c &= \expsub{\tilde{\Phi}}{\sum_{(X_i,G_i)\in\tilde{\Phi}}\ind{G_iR_i^{-\alpha}\geq \eta}} \nonumber \\
&= 2\pi\lambda\int_0^\infty\int_0^\infty \ind{GR^{-\alpha}\geq \eta}Re^{-G}\diff R \diff G \nonumber \\
&= \frac{2}{\alpha}\pi\lambda\eta^{-2/\alpha}\Gamma(\frac{2}{\alpha})
\end{align}
where $\ind{\cdot}$ is the indicator function and $\Gamma(\cdot)$ is the Gamma function.
\begin{comment}
Furthermore,
the average number of neighbors is related to the
threshold $\eta$ by $c = \len \, \Prob(GR^{-\alpha} > \eta)$,
\LZ{which can be further calculated as $\frac{2}{\alpha} \lambda \pi
  \eta^{-4/\alpha} \Gamma(\frac{2}{\alpha})$ in case of Rayleigh
  fading, where % $\lambda= \len/A$ is the node density of the network and
  $\Gamma(\cdot)$ is the Gamma function.}
\end{comment}

The near-far situation, namely that some neighbors can be much
stronger than others is inherently modeled
in~\eqref{eq:Y}-\eqref{eq:Un_pdf}.
The proposed sparse recovery algorithms are highly resilient to the near-far
problem. In particular, in the case of deterministic signatures, the gain of strong neighbors
can be estimated quite accurately so that their interference to weaker
neighbors can be removed.

\begin{comment}
In case of Rayleigh fading, %we have % use~\eqref{eq:Gt} to obtain
\begin{align}
 c & = \frac{N}{A} \pi \eta^{-2/\alpha}
 \int_0^{\infty}G^{2/\alpha}e^{-G}dG
  = \frac{2\pi N}{\alpha A} %\lambda \pi
 \eta^{-2/\alpha}
 \Gamma\left(\frac{2}{\alpha}\right)
\end{align}
where % $\lambda = \len / A$ is the node density and
$\Gamma(\cdot)$ is
the Gamma function.
\end{comment}

\subsection{Network-wide Discovery}
\label{s:netwide}

Unlike in previous work~\cite{LuoGuo08Allerton, LuoGuo09Allerton},
this paper also considers the problem that many or all nodes in the
network need to discover their respective neighborhoods at the same
time.  A major challenge is posed by the {\em
  half-duplex} constraint, i.e., that a wireless node cannot receive
any useful signal at the same time and over the same frequency band on
which it is transmitting~\cite{Rappap02, Ramana99WN}.  This is due to
the limited dynamic range of affordable radio frequency circuits.
Standard designs of wireless networks use time- or frequency-division duplex to
schedule transmissions of a node away from the time-frequency slots
the node employs for reception~\cite{XuSaa01CM}.

\begin{comment}
  It is important to note that a wireless node essentially cannot
  receive any useful signal at the same time and over the same
  frequency band within which it is transmitting~\cite{Rappap02,
    Ramana99WN}.  This is while it is transmitting over the same radio
  frequency.  This so-called {\em half-duplex} constraint is largely
  due to the limited dynamic range of affordable radio frequency
  circuits, which is likely to remain a physical restriction in the
  foreseeable future despite advances in technology.
\end{comment}

A random-access scheme naturally supports network-wide discovery.  This
is because each node transmits its NIA intermittently, so that it can
listen to the channel to collect neighbors' NIAs during its own
epochs of non-transmission.  Collision is inevitable, but if
each node repeats its NIA a
sufficient number of times with enough (random) spacing, then with
high probability it can be received by every neighbor once
without collision.

As we shall see in Sections~\ref{s:onoff} and~\ref{s:rm}, the
proposed compressed neighbor discovery schemes employ
on-off signatures, so that a node can make
observations during its own off-slots.
All nodes broadcast their signatures and discover their respective
neighbors at the same time.  Thus network-wide discovery is achieved within a single frame interval.

\section{Random Signatures and Group Testing}
\label{s:onoff}

In this section, we consider using random on-off
signatures.  Specifically, the measurement matrix $\SSS$ consists of
i.i.d.\ Bernoulli random variables, with $\Prob(S_{mn}=1) =
1-\Prob(S_{mn}=0) = q$ for all $m,n$.  As aforementioned, the
signatures can be generated using a common pseudo-random number
generator so that $\SSS$ is known to all query nodes.

\subsection{A Previous Algorithm Based on Group Testing}
\label{s:group}

In the absence of noise, neighbor discovery with on-off signatures is
equivalent to the classical problem of group testing.
In fact, group testing has been used to solve a related RFID
problem~\cite{KodLau09Wiopt} and a multiple access
problem~\cite{BerMeh84TC, KurSid88TC}.
For every $m=1,\dots,M$, the measurement
  $Y_\mmi = \sqrt{\snc} \sum_{\lei = 1}^\len  %{\lei \in \N_0} %
  S_{\mmi\lei} X_\lei$ is nonzero if any node from the group $\{\lei\,:\,
\lei=1,\dots,\len,$ and $S_{\mmi\lei}\ne0\}$ is a neighbor.
Algorithm~\ref{alg:Elimination}
%\DG{of}~\cite{LuoGuo09Allerton}
visits every
measurement $Y_\mmi$ with its power below a threshold $\thr$ to
eliminate all nodes who would have transmitted energy at time $\mmi$
from the neighbor list.  Those nodes which survive the elimination
process are regarded as neighbors.

\begin{algorithm} %[t]
  \caption{Simple group testing}\label{alg:Elimination}
  \begin{algorithmic}[1]
    \STATE {\bf Input:} $\Y$, $\SSS$ and $T$ %
    \STATE {\bf Initialize:} $V \leftarrow \{1,\dots,\len\}$ %
    \FOR{$i = 1$ to $\mmt$}
    \IF{$|Y_m|^2<T$}
    \STATE $V \leftarrow V \,\backslash\, \{\lei:\, S_{\mmi\lei}=1\}$    \ENDIF
 \ENDFOR
 \STATE {\bf Output:} mark all nodes in $V$ as neighbors
\end{algorithmic}
\end{algorithm}

\begin{comment}
\begin{figure}
  \centering
  % \includegraphics[width = 5in]{InfocomSimulationStrikes1}
  \includegraphics[width=\columnwidth]{../figure/InfocomSimulationStrikes1}
  \caption{The probabilities of error versus SNR.
    No false alarm is registered in the experiment.
    % ,    so $P_f$ is numerically zero.
    % The parameters are: $\len = 10,000$, $c = 5$,
    % $\mmt = 1,500$, $q = 0.0247$ and $\thr = 1.0$.
  }
  \label{f:PmPf}
\end{figure}
\end{comment}

Two types of errors are possible: If an actual neighbor is eliminated
by the algorithm, it is called a {\em miss}. On the other hand, if a
non-neighbor survives the algorithm and is thus declared a neighbor,
it is called a {\em false alarm}.
The {\em rate of miss} (resp.~{\em rate of false alarm}) is defined as
the average number of misses (resp.~false alarms) in one node's
neighborhood divided by the average number of neighbors a node has.

Algorithm~\ref{alg:Elimination} requires only noncoherent
energy detection and is remarkably simple.
However, discovery is reliably only if
the SNR is high and the average number of neighbors a node has is very
small, whereas the error performance is unacceptable for many practical
scenarios~\cite{LuoGuo09Allerton}.

\begin{comment}
From the viewpoint of a query node, the neighborhood output of the
algorithm is in error if there is any miss or false alarm.

In the following, we simulate the model~\eqref{eq:Y} and
Algorithm~\ref{alg:Elimination} under a fairly practical scenario.
For simplicity, let the signatures be i.i.d.~Bernoulli($q$) sequences.
Consider a system with $\len = 10,000$ nodes besides node 0.  Node $0$
has in average $c=5$ neighbors, hence the number of neighbors is a
binomial($\len,c/\len$) random variable.  Rayleigh fading is assumed and
the path loss exponent $\alpha=3$.  Let the signature length be $\mmt =
1,500$ and let $q = 0.0247$, so that each signature consists of about
37 pulses on average.  The algorithm uses the threshold $\thr=1$.
These parameters are optimized for a slightly different algorithm
(Algorithm~\ref{alg:Multiple} in Section~\ref{s:improved}), but are
also a good choice for Algorithm~\ref{alg:Elimination}.
Fig.~\ref{f:PmPf} plots the probabilities of miss and false alarm
against the SNR.
%In particular, no false alarms are registered so that
The false alarm rate is effectively zero and insensitive to the SNR.
% In fact, once the SNR exceeds a few dB,
The rate of miss decays with the SNR and drops to about 0.01 at
35 dB.
\end{comment}

\subsection{Improvements: $t$-Tolerance Test and Phase Randomization}
\label{s:improved}

The improved scheme proposed in this section is based on
Algorithm~\ref{alg:Elimination} and includes two major changes.
First, instead of eliminating a node as soon as it disagrees with one
measurement, multiple disagreements are needed to eliminate a node.  Secondly, to decouple the measurements, we randomize the phase
of the samples of each signature.

% With a closer look at the algorithm, the probability of false alarm is
% very good but the probability of miss is quite high for moderate SNRs.

\begin{figure}
  \centering
  \includegraphics[width=5in]{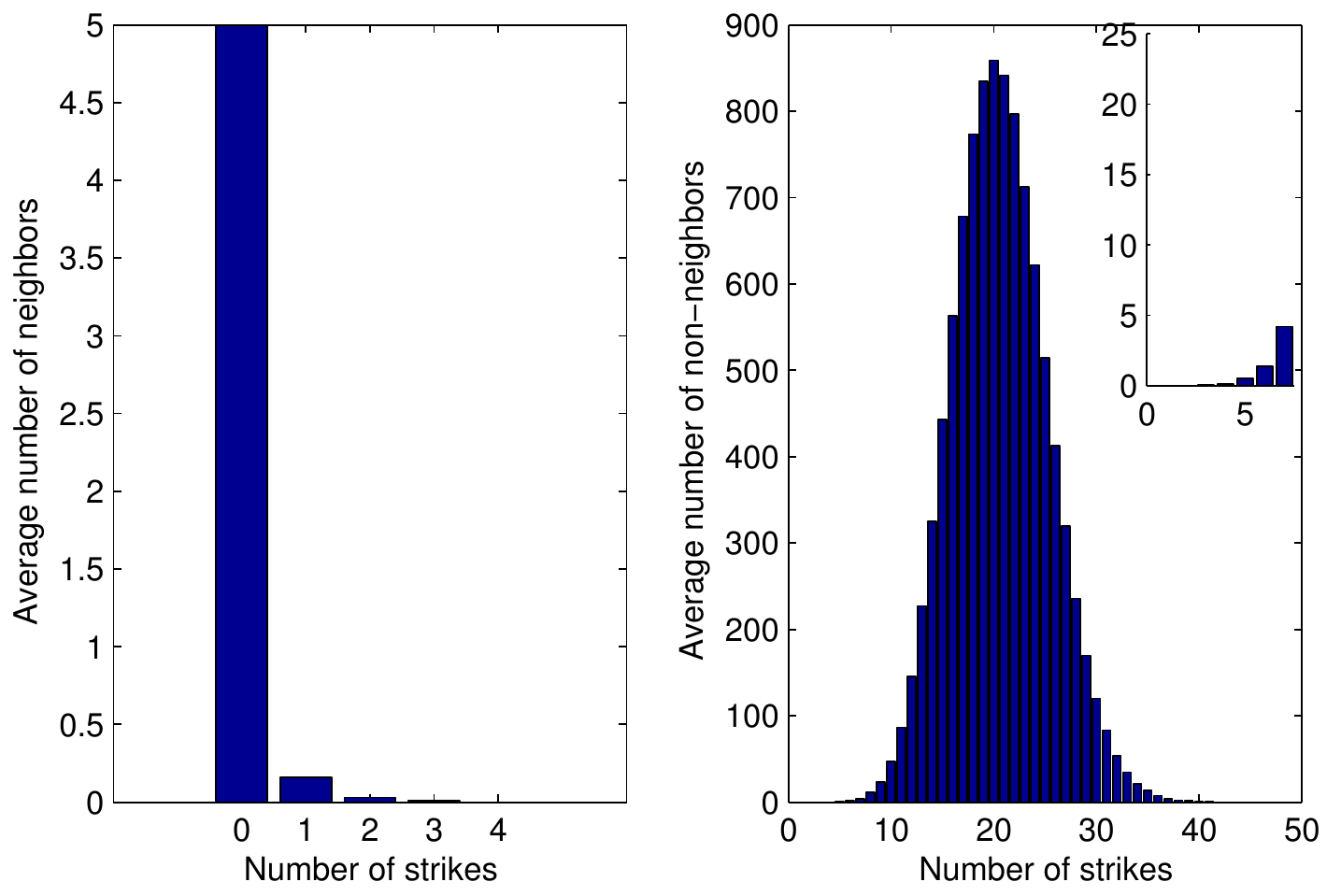}
  \caption{Histograms: The average number of
    nodes %(out of 10,000 total)
    versus the number of strikes they receive as a neighbor (the left
    plot) or non-neighbor (the right plot).  The inlet plot amplifies
    the left-side tail. \LZnew{SNR = 20~dB, $c=5$.}}
  \label{f:hist}
\end{figure}

It is instructive to examine the events which trigger elimination.  To
this end, we record, for each node eliminated, the number of near-zero
measurements which point to its elimination, which are referred to as
{\em strikes}.
\DG{Fig.~\ref{f:hist} illustrates the average number of nodes (out of
10,000 total) which receive 0,1,2,$\dots$ strikes as neighbors or
non-neighbors, respectively.
It turns out that most of the time a neighbor agrees with all
measurements and hence receives no strike, but occasionally a
neighbor may receive 1, 2 or 3 strikes due to noise or mutual
cancellation.   In contrast, most non-neighbors receive dozens of
strikes because they disagree with many measurements, whereas a
small number of non-neighbors receive fewer than 5 strikes. }
\begin{comment}
In other words, a neighbor who is eliminated receives mostly
one strike, and occasionally two strikes or more, whereas most non-neighbors
receive many more strikes.  The reason is that at moderate to high
SNR, two neighbors both receive a strike if their pulses cancel at
the same measurement, because they each transmit a pulse with
approximately opposite channel coefficient.  This is infrequent
because signatures are quite sparse.  On the other hand, a
non-neighbor gets struck many times because it would have transmitted
over many symbol intervals with zero measurement.
\end{comment}
Algorithm~\ref{alg:Multiple}, which is referred to as the $t$-tolerance
%$k$-strike
test~\cite{KniBru98DAM}, allows a node receiving up to $t$ strikes to survive, and
requires that a node be eliminated only if it receives strictly more
than $t$ strikes.
%at least $k$ strikes.
By tuning the number $t$, one can select the most desirable trade-off
between the rate of miss and the rate of false alarm.

\begin{algorithm}[h!]
  \caption{$t$-tolerance group testing}\label{alg:Multiple}
  \begin{algorithmic}[1]
    \STATE {\bf Input:} $\Y$, $\SSS$ and $T$ %
    \STATE {\bf Initialize:} $v_n \leftarrow t+1, n = 1, \dots, N$ %
    \FOR{$i = 1$ to $\mmt$}
    \IF{$|Y_m|^2<T$}
    \STATE $v_n \leftarrow v_n - S_{mn}, n = 1,\dots, N$   \ENDIF
 \ENDFOR
 \STATE {\bf Output:} $\{ n : v_n > 0, n = 1, \dots, N \}$
\end{algorithmic}
\end{algorithm}

We further examine
one of the major causes of misses, which is that the pulses of two
or more neighbors cancel at the receiver, so that the measurement
$Y_\mmi$ is below the threshold at multiple intervals.  This takes
place for two neighbors $n_1$ and $n_2$ if their channel
coefficients are similar in amplitude but opposite in phase, so
that $S_{\mmi\lei_1}U_{\lei_1} + S_{\mmi\lei_2}U_{\lei_2} \approx
0$ for every interval $\mmi$ where both nodes transmit a pulse,
which implies the neighbors will be eliminated erroneously with a
number of strikes wherever their pulses coincide.

A simple trick can be used to reduce misses with essentially no impact
on false alarm.  The idea is to let each node randomize the phases of
its signature at different slots independently, i.e., use
$S_{\mmi\lei} e^{j\Theta_{\mmi\lei}}$ in lieu of $S_{\mmi\lei}$ where
$\Theta_{\mmi\lei}$ are i.i.d.~uniform on $[0,2\pi)$.  In this case,
if $S_{\mmi\lei_1}e^{j\Theta_{\mmi\lei_1}}U_{\lei_1} +
S_{\mmi\lei_2}e^{j\Theta_{\mmi\lei_2}}U_{\lei_2} \approx 0$ for some
slot $\mmi$, it is unlikely that this is still true for other slots.
We note that the randomization is easy to implement at transmitters
and requires no
change at the receivers, because knowledge of the phases is
not needed by the noncoherent detection algorithm.

\begin{comment}
Similar analysis as in Appendices~\ref{app:falseMultiple}
and~\ref{app:UpperBoundMiss} suggest that the preceding optimized
designs are still valid when signatures have random phases.

Precisely, we modify the element-wise system model~\eqref{eq:NDdesign}
so that it reads as
\begin{equation}\label{eq:NDdesign}
  Y_\mmi = \sqrt{\snc} \sum_{\lei = 1}^\len S_{\mmi\lei} e^{-j \phs_{\mmi \lei}}
  U_\lei B_\lei
  + W_\mmi,\ 1\leq \mmi \leq \mmt,
\end{equation}
where $\{\phs_{\mmi\lei}\}_{\mmi = 1,\dots, \mmt; \lei =
1,\dots,\len}$ are i.i.d. uniform random variables over the
interval $[0, 2\pi)$. The intuition behind the use of random phase
is that it reduces the chance that two neighbors cancel each other
at multiple chips such that the probability of miss is reduced.
\end{comment}

% We note that false alarms can be more detrimental than misses to the
% stability and fidelity of wireless networks,

\subsection{Design Optimization}
\label{s:opt}

The simple group testing algorithm and the improved $t$-tolerance
algorithm are in general difficult to analyze.
In the absence of noise, there have been asymptotic results (see,
e.g., \cite{BerMeh84TC, LuoGuo08Allerton}) where it is shown that the
error probabilities vanish as the problem size increases as long as
the number of measurements exceed a certain level, which depends
typically logarithmically on the node population.
Under a discrete model with noisy measurements, asymptotic performance
bounds for the algorithm have been developed in~\cite{KniBru98DAM}.
Other studies of the theoretical limits of noisy measurement models
with 0-tolerance test~\cite{CheHor09Allerton, CheKar10ISIT, Che10Allerton,
   AtiSal10X} assume the active signatures to be transmitted at equal
 power, and thus do not apply to the current model~\eqref{eq:Y} with
 fading and path loss.

%Unknown fading and additive Gaussian noise further complicates the
%analysis.
No existing analytical results or techniques for group testing yields
a good approximation of the performance of
Algorithm~\ref{alg:Multiple}.  Therefore, we resort to numerical
methods to find the optimal design trade-off between cost and
error performance.  The cost here refers to the signature length
(i.e., the neighbor discovery overhead) and the SNR.
We assume that the node population and density, the SNR, as well as
the fading characteristics are given, which are not controlled
by the designer.

For a given signature length $M$, a good indicator of the rate of miss
is $\mmt q$, i.e., the average number of active pulses in the
signature.  Intuitively, for an actual neighbor, the larger $\mmt q$ is,
the more likely its symbols get canceled by other nodes, thus causing
unwanted strikes and misses.  On the other hand, if $q$ is too small,
some non-neighbors may not receive enough strikes to be eliminated,
thereby causing false alarms.  Similar qualitative statements can be
made about the threshold $T$ used in Algorithms~\ref{alg:Elimination}
and~\ref{alg:Multiple}.  It is not difficult to see that $T$ should be
above the noise variance, but perhaps not too much higher than that.

In this work, we assume the signature length $M$ is given and fixed.
We then numerically search the optimal choice of the sparsity $q$ and
threshold $T$ as those that minimize the total rate of miss and false
alarm.
Since using identical signatures under all channel conditions is
preferable in practice, the numerical search is carried out under a
specific SNR.  The same parameters $q$ and $T$ are then used at all
other SNRs.

\subsection{Numerical Results}\label{s:optimal}

We next show some numerical results obtained based on the
design described in Section~\ref{s:opt}.

% design: The optimal sparsity and threshold, $(q, T)$, can be obtained
% at each SNR according to~\eqref{eq:minq}.
% %\DG{The are then fine-tuned based on a small number of numerical experiments.}
% %It turns out that the optimal values of $q$ do not differ much for a wide range of SNR.
%  For each SNR, we then fine tune the threshold $T$ used by the
%  receiver numerically, such that the error rate is minimized.

\begin{figure}[t]
  \centering
  \includegraphics[width=5in]{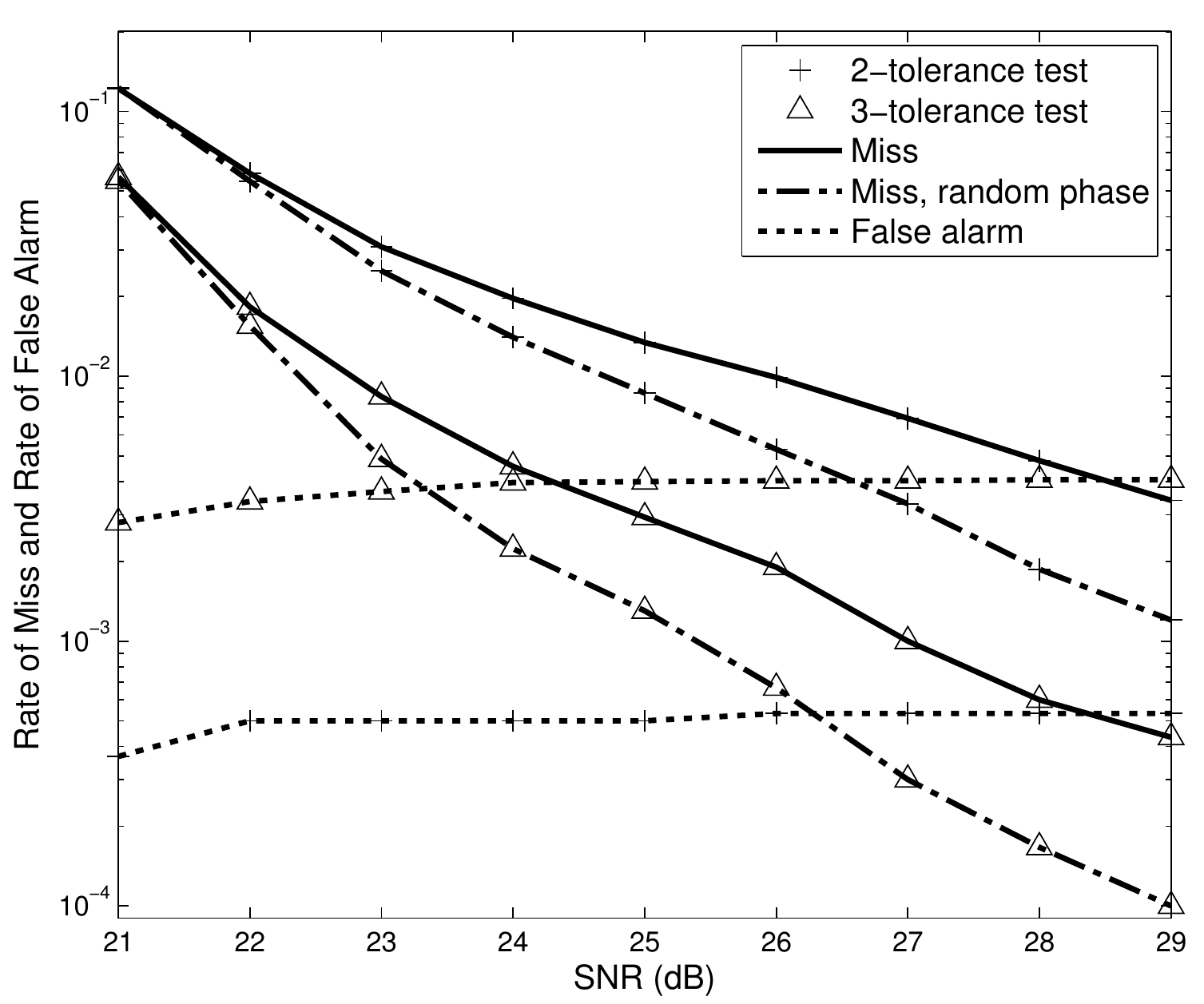}
  \caption{Rates of miss and false alarm versus SNR.  In all 1,000
    trials, $\len=1$0,000, $c=3$0, $\mmt=2$,048, and $q=0.0176$. }
\label{f:Pe30}
\end{figure}

% Figs.~\ref{f:Pe50} and~\ref{f:Pe10} plots numerical results
%Consider first a network with $\len = 1$0,000 nodes besides node 0,

Suppose there are $\len=1$0,000 valid NIAs
which belong to nodes uniformly distributed in a square centered
at the origin. Let the path loss exponent be $\alpha=3$.  Assume Rayleigh fading and that a node is regarded as a neighbor if the channel gain exceeds
$\eta=0.05$.  In each network realization, we consider the average neighbor discovery performance of the 100 nearest nodes to the origin.

Let the density of the network be that there are
on average $c =3$0 nodes in each neighborhood.
%, Rayleigh fading and path loss exponent $\alpha = 3$.
Each signature consists of $M=2$,048
symbols.  At 28 dB SNR, the optimal sparsity and threshold are found
to be $q=0.0176$ and $T=2.0$, respectively, for a 2-tolerance test.
The random signature matrix is generated with this fixed sparsity, so
that there are on average $Mq=36$ pulses in a signature.  The same
signature matrices and threshold are then used at all SNRs in all tests.
% (* 2048 .0176)

The receiver carries out Algorithm~\ref{alg:Multiple} and the
resulting rates of miss and false alarm are plotted against the SNR in
Fig.~\ref{f:Pe30}.
The rate of false alarm is plotted in dotted lines, the rates of miss
with and without random-phase improvement are plotted in dash-dotted
lines and solid lines, respectively.  The performance of the
2-tolerance test is marked with '+' and that of the 3-tolerance test
is marked with '$\Delta$.'

In case of the 2-tolerance test, missed neighbors are the dominant source
of error.  Using random phases improves the rate of miss significantly.
The rate of miss decreases with the SNR and drops to 0.1\% at 29 dB
(with random phases).
The rate of false alarm is not sensitive to the SNR and stays around 0.05\%.
Using the 3-tolerance test improves the rate of miss significantly (about
4 dB at the error rate of 0.1\%), because actual neighbors are less
likely to be eliminated.  The rate of false alarm, however, becomes
higher with higher tolerance.  If the total error rate is of concern,
then the 2-tolerance test is preferable if the SNR is above 27
dB, whereas the 3-tolerance test is preferable otherwise.

\begin{comment}
Although the rate of false alarm increases, the total error rate
is less than 1\% at 23 dB or higher.
The error probabilities achieved by Algorithm~\ref{alg:Multiple} with
$k=2$ and phase randomization are plotted against the SNR in
Fig.~\ref{f:Pe} (see the curves with the `$\square$' markers).  It is
clear that the probability of false alarm increases, whereas the
probability of miss decreases compared to the case without phase
randomization.  At 26 dB, the probability of error almost as low as
$10^{-3}$.
\end{comment}

\begin{figure}[t]
  \centering
  \includegraphics[width=5in]{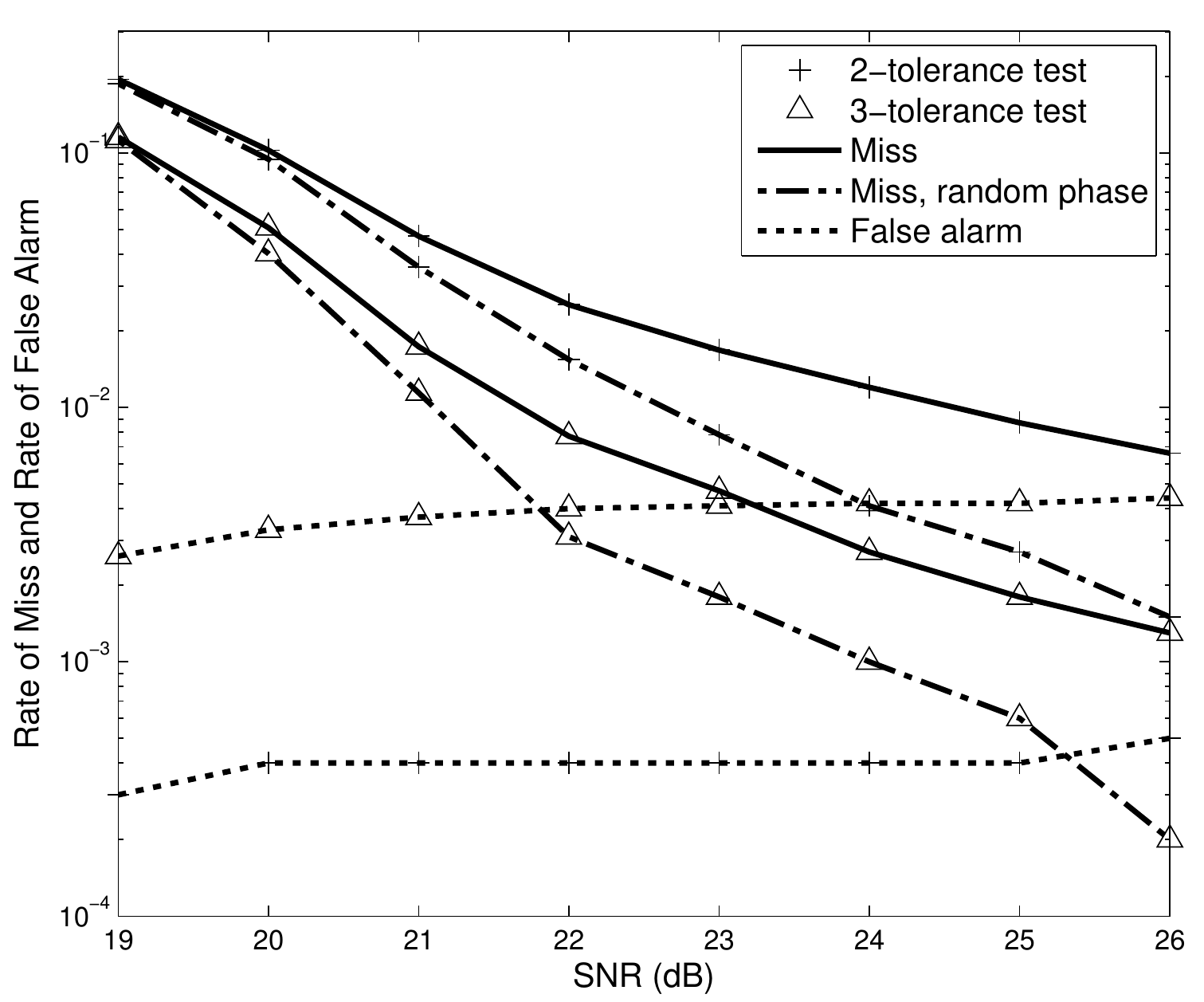}
 \caption{Rates of miss and false alarm versus SNR.  In all 1,000
    trials, $\len=1$0,000, $c = 1$0, $\mmt = 1$,024, and $q=0.0371$. }
\label{f:Pe10}
\end{figure}

\begin{comment}
Under the same situations and using identical design parameters, the
rate of miss achieved with phase randomization is also plotted in
Fig.~\ref{f:Pe50}, using dash-dot curves.
Clearly, the performance of both the 2-strike and 3-strike cases are improved slightly.
\end{comment}

Fig.~\ref{f:Pe10} repeats the preceding experiment, except for a
sparser network, where the average number of neighbors a node has is
$c=1$0, out of $\len=1$0,000 nodes, and that shorter signatures are
used: $\mmt=1$,024.  The parameters $q=0.0371$ and $T=3.0$ are
optimized for 26 dB for the 2-tolerance test and then used at all SNRs
and tests.  On average 38 pulses are found in a signature.
%(* 1024 .0371)

Fig.~\ref{f:Pe10} shows that using the 3-tolerance test yields better
error rates at high SNRs.  At 26 dB, the total error rate is just below 0.1\%.

It appears that the improvement from using phase randomization is
more pronounced with higher tolerance.  In case of 2-tolerance test,
the rate of miss can be reduced by about 10 fold at high SNRs.
%
\begin{comment}
At %SNR$=
$19$ dB, the total error rate can be as low as $0.6$\%,
achieved by using the 4-strike algorithm.  If the 3-strike algorithm
is used, the error rate is as low as $0.2$\% at SNR$=23$ dB.
\end{comment}

\subsection{Network-wide Neighbor Discovery}

Although the preceding development assumes a single query node in the
network, it is easy to extend the algorithms
to network-wide neighbor discovery, where all or any subset of nodes
acquire their neighborhoods simultaneously.
This is an advantage of using on-off signatures, because a node can
receive useful signal during its own off-slots despite of the
half-duplex constraint.
In fact the signatures are often very sparse
(e.g., $q<0.04$ in the preceding numerical examples), so that
``erased'' received symbols due to one's own transmission are few.
This also implies that even if the energy of a pulse leaks into
neighboring symbol intervals, there are still enough off-slots for
making observations.

The impact of the half-duplex constraint is in effect a reduction of
the length of the signatures.
From the viewpoint of any query node, once the erasures are purged,
models~\eqref{eq:Y} and~\eqref{eq:YS} still apply, if the number of
measurements $\mmt$ is replaced by a random variable of binomial
distribution with parameters ($\mmt,1-q$).  For large $\mmt$, the
number of useful measurements is approximately $\mmt(1-q)$.  The
discovery algorithm can be carried out by all nodes simultaneously.
% The probability of false alarm can be calculated in the same manner as
% in the proof of Proposition~\ref{pr:Pf}, but requires averaging over
% the binomial distribution.  %The result is given in Appendix ?.
If we increase $\mmt$ by a factor of $1/(1-q)$, then the performance
of network-wide neighbor discovery is roughly the same as in the case
of a single query node with the original signature length.

\begin{comment}
  For example, node $0$ can recover its neighborhood by
solving the compressed sensing problem described by~\eqref{eq:YS} with
all rows of $\SSS$ corresponding to nonzero entries of its own
signature $\S_0$ struck out, and the dimensions of the signals
adjusted accordingly.  In fact all nodes can carry out neighbor
discovery similarly at the same time.  As we shall see, this much
relaxed structure of the signatures yields significant performance
improvement over random access discovery.
\end{comment}

\subsection{Computational Complexity}
\label{s:complexity}

\DG{
After turning the measurements $\Y$ into a binary $N$-vector by
comparing it with a threshold, all computations carried out by
Algorithms~\ref{alg:Elimination} and~\ref{alg:Multiple}  are
binary or counting down by 1.
%The group testing algorithms visit the columns of the matrix $\SSS$
%corresponding to nonzero entries of $\hat{\Y}$.
The computational
complexity is $\mathcal{O}(N M \,q)$} \LZnew{if implemented in a clever way using the sparsity of the signature matrix. If network-wide neighbor discovery is carried out, the complexity at each decoder is increased by a factor of $1/(1-q) \approx 1+q$, since $q$ is typically a very small number.}
A general purpose processor may handle up to $\len=10^5$ NIAs in real
time (where $\mmt$ is typically a few thousand).  Hardware
implementation using, for example, a programmable gate array, may take
advantage of the fact that the elimination procedure
% (which is similar to group testing)
can be carried out in parallel. In this case, it is
conceivable to carry out compressed neighbor discovery for a large
address space including all 32-bit Internet Protocol (IP) addresses.

An alternative, more scalable approach proposed in~\cite{LuoGuo08Allerton}
is to divide the address into smaller segments
(e.g., a 32-bit address consists of three overlapping 16-bit
subaddresses), and discover the subaddresses of all neighbors
separately using the preceding algorithms.  The subaddresses can then
be pieced together to form full addresses by matching their overlaps.

\DG{
A natural question to ask is why noncoherent group testing algorithms
are proposed in this paper in lieu of coherent detection, such as
matched filtering followed by thresholding,
which should perform better.  The reason is that even simple matched filtering
entails a much higher complexity with $\mathcal{O}(NM)$ additions over
the precision of the measurements.
}

%We note that
The problem of inferring about the inputs to a noisy
linear system from the outputs have been studied in many contexts.
One important area relevant to the model~\eqref{eq:YS} is multiuser
detection.  References~\cite{LinLim04TC, AngBig07ICASSP}
considers a related user activity detection problem in cellular
networks, and suggest the use of coherent multiuser detection
techniques.  Such techniques do not apply here because they require
knowledge of the channel coefficients $U_\lei$ of all neighbors, which
is clearly unavailable before the neighbors are even known.
Reference~\cite{AngBig10PhyCom} considers channel estimation, but the
algorithm is more complex than matched filtering, and thus does not
scale well with the network size.

The idea of using a $t$-tolerance test in Algorithm~\ref{alg:Multiple}
is related to the wisdom
of {\em belief propagation}, where the decision for each node at
question is made using beliefs provided by all relevant measurements.
One can in fact carry out belief propagation fully and
iteratively~\cite{KscFre01IT}, but we suspect the performance gain
does not justify the additional complexity here.
%, which allows one to eliminate nodes
% which disagree with the most number of measurements.

% Specifically, the $\mmt\times\len$ binary measurement matrix $\SSS$
% can be stored in the array.  Once the $\mmt$ measurements are
% converted to $\mmt$ bits after thresholding, one can go through all
% the zeros, where for each zero measurement, simultaneously eliminate
% all incompatible nodes (signatures).

\begin{comment}
\DG{If network-wide neighbor discovery is carried out,
a fraction $q$ of the measurements are lost due to self interference.
To compensate for this loss, the number of measurements should be
increased by a factor of $1/(1-q)$, which is small since $q$ is
typically a very small number.
The complexity of group testing becomes $\mathcal{O}(M N
c\,q/(1-q))$.}
%, which is roughly the same as $\mathcal{O}(\mmt\len q)$ since $q$ is
%a very small number.
\end{comment}

\DG{
As long as random signatures are used, any good decoding algorithm
needs to visit every signature, so that the complexity is at least linear
in the address space $N$.  This prohibits scaling to a very large
space, say $N=2^{48}$.  Although random signatures perform
as good as any signatures according to Shannon's random
coding argument, it is well-known that structures need to be
introduced in the codebook in order for low-complexity decoding.
This is the subject of the next section.
}

\renewcommand{\B}[1]{\boldsymbol{#1}}

\section{On-off Reed-Muller Signatures and Chirp Decoding
}\label{s:rm}

In this section, we propose to use deterministic signatures
obtained from second-order Reed-Muller codes with erasures,
where the complexity of the corresponding chirp decoding algorithm
is sub-linear in $N$.
We first discuss the original RM code without erasure.  Such a code is
sufficient for a single silent query node to acquire its neighborhood.
The construction of the RM code is described in detail in~\cite{CalGil06X}.
We provide a sketch of the construction in Section~\ref{s:rm1}.
The signatures consist of QPSK entries, which prevent a transmitting
node from simultaneously discovering its neighborhood.
In Section~\ref{s:rm0}, zero entries are introduced by erasing about
$50\%$ of the symbols in each signature, so that full-duplex neighbor
discovery is enabled.  The chirp decoding algorithm is discussed in
Section~\ref{s:chirp}.  As we shall see in Section~\ref{s:example},
using the Reed-Muller code enables more reliable and
efficient discovery in networks which are many orders of magnitude
larger than allowed by using random on-off signatures.

\DG{For the reader's convenience, the signature generation and chirp
  decoding procedures are summarized as Algorithms 3 and 4.  Examples
  in the case of very small systems are given to illustrate the
  encoding and decoding procedures.}

%DG: (cf.~Section~\ref{s:complexity}).
\begin{comment}
The key to this scheme is to first
construct a deterministic measurement matrix based on \DG{a} %the
second order
Reed-Muller code. %s.

We first discuss the original RM code without erasure.  Such a code is
sufficient for a single silent query node to acquire its
neighborhood. The construction of the signatures is described in
detail in~\cite{CalGil06X}.
We %only
provide a sketch of the construction.
\end{comment}

\subsection{The Reed-Muller Code (without Erasure)}
\label{s:rm1}

RM codes are a family of linear error-control codes.  A formal
description of RM codes requires a substantial amount of preparation
in finite fields.
In a general form, RM codes are based on evaluating certain primitive
polynomials in finite fields.
Due to space limitations, we briefly describe the second-order RM
codes used in this paper using the minimum amount of formalisms.
The reader is referred to~\cite{CalGil06X} for a more detailed discussion.

Given a positive integer $m$, we show how to generate up to
$2^{m(m+3)/2}$ distinct codewords, each of length $2^m$.  For example,
in the case of $m=10$, there are up to $2^{65}$ codewords of length
1,024.

\begin{comment}
% $m$,
\DG{We first}
 construct the Kerdock set
$\B{\mathcal{K}}(m)$ consisting of $2^m$ binary symmetric $m\times m$
matrices.
%Roughly speaking,
They are binary Hankel matrices where the
top row consists of arbitrary entries and each of the remaining
reverse diagonals is computed from a fixed linear combination of the
entries in the top row.  It is known that
$\B{\mathcal{K}}(m)$ is an
\end{comment}

\LZnew{
Let $e_l^i=(0,\dots,0,1,0,\dots,0)$ be a row vector of length $l$ in
which the $i$-th entry is equal to $1$ whereas all other entries are
zeros. Let $\B{P}(e_l^i)$ be the $l\times l$ symmetric matrix in which
the top row is $e_l^i$ and each of the remaining reverse diagonals
(a diagonal from upper right to lower left) is computed from a fixed
linear combination of the entries in the top row.  The reader is
referred to~\cite{CalGil06X} for a detailed description of the
construction, which is based on evaluating some primitive polynomials
in GF($2^m$).
% How to choose the linear combination is clearly stated in [29]. Basically it is constructed
% upon the primitive polynomial in GF(2^m). I think we just need to refer to the paper [29].
% The reverse diagonals are those from upper right to lower left.
For example, $\B{P}(e_1^1)=1$ and
 \begin{align}
   \label{eq:Pe2}
   \B{P}(e^1_2) =
   \begin{bmatrix}
     1 \;\; 0 \\
     0 \;\; 1
   \end{bmatrix}, \;
   \B{P}(e^2_2) =
   \begin{bmatrix}
     0 \;\; 1 \\
     1 \;\; 1
   \end{bmatrix}.
 \end{align}

Given $m$, we form a linear space of $m\times m$ symmetric matrices with a set $\B{B}$ of $m(m+1)/2$ bases constructed using $\left\{\B{P}(e_l^i),i\leq l, l=1,\dots,m \right\}$, where for $l<m$, $P(e_l^i)$ is padded to an $m\times m$ matrix, where the lower right $l\times l$ submatrix is $\B{P}(e_l^i)$ and all remaining entries are zeros. In the simple case of $m=2$, $\B{B}$ consists of $m(m+1)/2=3$ bases, which are $\B{P}(e_2^1)$, $\B{P}(e_2^2)$ given by~\eqref{eq:Pe2} and an additional matrix obtained from $\B{P}(e_1^1)=1$ by padding zeros:
 \begin{align}
   \label{eq:e2}
   \B{B} = \left\{
   \begin{bmatrix}
     1 \;\; 0 \\
     0 \;\; 1
   \end{bmatrix}, \;
   \begin{bmatrix}
     0 \;\; 1 \\
     1 \;\; 1
   \end{bmatrix},  \;
   \begin{bmatrix}
     0 \;\; 0 \\
     0 \;\; 1
   \end{bmatrix}
   \right\}.
 \end{align}
Let $\B{B}(i)$ denote the $i$-th basis in $\B{B}$ ordered as $\B{P}(e_m^1),\dots,\B{P}(e_m^m)$ and then those obtained from $\B{P}(e_{m-1}^1),\dots,\B{P}(e_{m-1}^{m-1})$ and so on.}

Let the NIA consist of $n=n_1+n_2$ bits, where $n_1\le m$ and
$n_2\le m(m+1)/2$.  Each $n$-bit NIA is divided into two binary
vectors: $\B{b}' \in \mathbb{Z}_2^{n_1}$ and $\B{c} \in
\mathbb{Z}_2^{n_2}$,
\DG{where $\mathbb{Z}_2=\{0,1\}$.}
%with $n_1\le m$ and $n_2\le m(m+1)/2$.
 Let $\B{b}
\in \mathbb{Z}_2^m$ be formed by appending $m-n_1$ zeros after
$\B{b}'$ \DG{($\B{b}=\B{b}'$ if $n_1=m$)}.
\begin{comment}
Each node %can
then map\DG{s} its \DG{$n$-bit} NIA into a signature \DG{of QPSK
 symbols} as follows.  Let the
NIA consist of \DG{$n=$}$n_1+n_2$ bits, which is divided into two binary
vectors: $\B{b}' \in \mathbb{Z}_2^{n_1}$ and $\B{c} \in
\mathbb{Z}_2^{n_2}$, with $n_1\le m$ and $n_2\le m(m+1)/2$.
\end{comment}
\DG{We map $\B{c}$ to an $m\times m$ symmetric matrix according to
  \begin{align}
    \label{eq:3}
 \B{P}(\B{c})=\sum_{i=1}^{n_2} c_i\B{B}(i) \mod 2
 \end{align}
}%
 where $c_i$ denotes the $i$-th bit of $\B{c}$.
The corresponding codeword is of $2^m$ symbols, whose entry indexed by
$\B{a} \in \mathbb{Z}_2^m$ is given by
\begin{equation}\label{eq:RMFun}
 \phi_{\B{b},\B{c}}(\B{a}) = \expb{j\pi\left(\frac12
 \B{a}^{\textsf{T}}\B{P}(\B{c})\B{a} + \B{b}^{\textsf{T}}\B{a} \right)} \ .
\end{equation}
%The system can thus accommodate up to $2^{m(m+3)/2}$ nodes with
%distinct signatures, each of length $2^m$.
For example, in case $m=2$, there are up to $2^{m(m+3)/2}=32$
  codewords of length $2^m=4$.  Moreover, if the number of nodes is \LZnew{$16$, i.e., $n=4$}, %$n=2^4=16$,
  only 16 codewords are generated as functions of $(\B{b},\B{c})$ and
  given as column vectors in Table~\ref{t:RMcode}, where only the
  first two bases in~\eqref{eq:e2} are used as $n_1=n_2=2$.

\begin{table}[h]
\caption{16 Reed-Muller codewords.}
\begin{center}
%\begin{tabular}{|cccccccccccccccc|}
\setlength{\tabcolsep}{5pt}
%\begin{tabular}{|r|r|r|r|rrrrrrrrrrrrr|}
\begin{tabular}{|c|r|r|r|r|r|r|r|r|r|r|r|r|r|r|r|r|}
\hline
$\B{b}$ & $0$ & $1$ & $2$ & $3$ &
$0$ & $1$ & $2$ & $3$ &
$0$ & $1$ & $2$ & $3$ &
$0$ & $1$ & $2$ & $3$ \\
\hline
$\B{c}$ & $0$ &$0$&$0$&$0$&
$1$ &$1$&$1$&$1$&
$2$ &$2$&$2$&$2$&
$3$ &$3$&$3$&$3$\\
\hline
&$1$&$1$&$1$&$1$&   $1$&$1$&$1$&$1$&   $1$&$1$&$1$&$1$&   $1$&$1$&$1$&$1$ \\
\LZ{$\phi_{\B{b},\B{c}}$} %_{\B{P}(\B{c}),\B{b}}$
&$1$
&-$1$
&$1$
&-$1$
&$j$
&-$j$
&$j$
&-$j$
&$j$
&-$j$
&$j$
&-$j$
&$1$
&-$1$
&$1$
&-$1$
\\
&$1$
&$1$
&-$1$
&-$1$
& $1$
& $1$
& -$1$
& -$1$
& $j$
& $j$
& -$j$
& -$j$
& $1$
& $j$
& -$1$
& -$j$
\\
&$1$
&-$1$
&-$1$
&$1$
&-$j$
&$j$
&$j$
&-$j$
&-$1$
&$1$
&$1$
&-$1$
&-$1$
&$j$
&$1$
&-$j$
\\
\hline
\end{tabular}
\end{center}
\label{t:RMcode}
\end{table}

\subsection{Generation of On-Off Signatures}
\label{s:rm0}

The drawback of using the original RM code is that the codewords
defined by~\eqref{eq:RMFun} consist of QPSK symbols, so that a node
cannot simultaneously receive useful signals while transmitting its own
codeword.  In order to achieve full-duplex neighbor discovery, we
propose to erase about 50\% of the entries of each
codeword to obtain an on-off signature, so that nodes can listen
during their own off-slots. The signature of each node consists of
roughly as many off-slots as on-slots, thus two nodes can receive
pulses from each other over about 25\% of the slots.

For reasons to be explained shortly in conjunction with the chirp
decoding algorithm, we apply random erasures
to the signatures in the following simple manner:  Suppose $n_2$ is
chosen such that the $m\times m$ symmetric
matrix generated by each node is
determined by its first $m_0\leq m/2$ rows. For node $k$, the erasure
pattern $\B{r}_k$ of length $2^m$ is constructed as follows: Divide
$\B{r}_k$ into $2^{m_0}$ segments with equal length $2^{m-m_0}$, let
the first segment consist of i.i.d.\ Bernoulli random variables with parameter
$1/2$ and all remaining segments be identical copies of the first
segment. It is easy to see that after introducing erasures in %to
the
signatures, the network can still accommodate
%$2^{\frac{1}{8}m(3m+10)}$
$2^{m(3m+10)/8}$
nodes. For example, if $m=10$, %12$,
we have up to
$2^{\DG{50}}$ %69}$
signatures of length 1,024. %$2^{12}=4,096$.

\DG{The procedure for generating the on-off signatures based on the RM
  code is
%The signature generation for each node in the network is
summarized as Algorithm~\ref{alg:SigGen}.
}

\begin{algorithm}[h]
\caption{Signature Generation Algorithm}
\label{alg:SigGen}
\begin{algorithmic}[1]
\DG{
\STATE {\it Input:} $n$-bit NIA
% [As a trivial example just for illustration, let $n=2^4=16$.]
\STATE Choose $m$ such that $n=n_1+n_2$ with $n_1\leq m$ and $n_2\leq
\frac{m_0}{2}(2m-m_0+1)$ where $m_0\leq m/2$.
% [In the example, $n_1=2, n_2=2, m=2, m_0=1$.]
\STATE Divide $n$-bit NIA into two vectors $\B{b}' \in
\mathbb{Z}_2^{n_1}$ and $\B{c} \in \mathbb{Z}_2^{n_2}$.
Form $\B{b} \in \mathbb{Z}_2^m$ by appending $m-n_1$ zeros after $\B{b}'$.
% Let $\B{b} \in \mathbb{Z}_2^m$ be formed by appending $m-n_1$ zeros after $\B{b}'$.
\STATE Generate the original RM code \LZ{$\phi_{\B{b},\B{c}}$} of
length $2^m$ according to~\eqref{eq:RMFun}.
% [In the example, there are $16$ codewords listed in the following table.]
%\newline \\
%\vspace{0.5em}
\STATE Generate the erasure pattern $\B{r}$ of length $2^m$ as follows: Let the first segment of $2^{m-m_0}$ bits be i.i.d.\ Bernoulli random variables with parameter $1/2$ and repeat the segment $2^{m_0}$ times to form the $2^m$ bits of $\r$.
\STATE {\it Output:} The on-off signature of length $2^m$ is the
% pointwise
element-wise product of \LZ{$\phi_{\B{b},\B{c}}$} and $\B{r}$.
}
\end{algorithmic}
\end{algorithm}

\subsection{The Chirp Decoding Algorithm}
\label{s:chirp}

We recall that each node makes observations
via the multiaccess channel~\eqref{eq:Y}, which is a superposition of
its neighbors' signatures subject to fading and noise.
An iterative chirp decoding algorithm has been developed
  in~\cite{HowCal08CISS} to identify the codewords of the RM code
  based on their noisy superposition.
The general idea is to take the Hadamard transform of the
  auto-correlation of the signal in each iteration
  to expose the coefficient of the
  digital chirps and then cancel the discovered
  signatures from the signal.

In case of full-duplex discovery, the original chirp decoding
algorithm with some modifications can be applied here for any node
(say, node 0) to recover its neighborhood based on
the observations  through its own off-slots (denoted as $\tilde{\Y}$).
\DG{The details are provided in Algorithm~\ref{alg:Chirp}.}

\begin{algorithm}[htp]
\caption{The chirp decoding algorithm}
\label{alg:Chirp}
\begin{algorithmic}[1]
\STATE {\it Input:} received signal $\Y$ in~\eqref{eq:YS}, signatures
of all other nodes $\SSS$ and its own erasure pattern $\B{r}$.
\STATE Choose %three parameters:
the maximum iteration number $T_{\max}$, the %power
threshold $\eta_0$ and the maximum number $n_0$ of weak nodes
discovered till termination.
\STATE Initialize the residual signal $\Y_r$ to
% $\tilde{\Y}$, which is
the pointwise product of $\Y$ and $1-\B{r}$.
\STATE Initialize the iteration number $t$ to 0, the neighbor set
$\N=\emptyset$ and the coefficient vector $\C=\emptyset$.
\STATE {\it Main iterations:}
\WHILE{$t \leq T_{\max}$}
    \FOR{$i=1,2,\dots,m_0$}
        \STATE Compute the pointwise multiplication of \LZnew{the
          conjugate of $\Y_r$ and the shift of $\Y_r$ in the amount of
          $2^{m-i}$.}
        %$\Y_r$ with its shift in the amount of $2^{m-i}$.
        \STATE Compute the fast Walsh-Hadamard transform of the computed auto-correlation.
        \STATE Find the position of the highest peak in the frequency domain
        %. Based on the peak location,
        and decode the $i$-th row of an $m\times m$ matrix $\B{P}(\B{c}_k)$, which corresponds to a certain node $k$.
    \ENDFOR
    \STATE Use the first $m_0$ rows of the preceding $\B{P}(\B{c}_k)$
    to determine its remaining rows.
    \STATE Compute $\S_k^0(\B{a})=\expb{j\pi\left(\frac12 \B{a}\tran{}\B{P}(\B{c}_k)\B{a} \right)}$ for all $\B{a} \in \mathbb{Z}_2^m$ and apply Hadamard transform to the pointwise product of $\Y_r$ and \LZnew{the conjugate of $\S_k^0$};
    \STATE Recover $\B{b}_k$ by finding the highest peak in the frequency domain.
    \STATE Compute $\phi_{\B{b}_k,\B{c}_k}$ according
    to~\eqref{eq:RMFun} and recover
    %the on-off signature
    $\S_k$ by pointwise product of $\phi_{\B{b}_k,\B{c}_k}$ and $\B{r}_k$.
    \STATE Add node $k$ to the neighbor set $\N$ and add a corresponding 0 to the coefficient vector $\C$.
    \STATE Put together all signatures of nodes in $\N$ to form a matrix $\SSS_{\N}$. Construct $\tilde{\SSS}$ by pointwise multiplying each column in $\SSS_{\N}$ with $1-\r$.
    \STATE Determine the value of vector $\X$ which minimizes $\|\Y_r-\tilde{\SSS}\X\|_2$. Update the coefficient vector $\C$ by $\C+\X$.
    \STATE Update the residual signal $\Y_r$ by $\Y_r-\tilde{\SSS}\LZnew{\X}$.
    \IF{\LZ{$\N$ contains more than $n_0$ nodes with coefficients less than $\eta_0$}}
        \STATE \LZ{Stop the main iteration.}
    \ENDIF
\ENDWHILE
\STATE {\it Output:} All elements in $\N$ whose
corresponding coefficients in $\C$ are no less than $\eta_0$.
\end{algorithmic}
\end{algorithm}

\DG{In the following, we provide a simple example to illustrate the
  key steps of Algorithm 4.
Consider a network of $N=2^n=\,$1,024 nodes.
Let the parameters in Algorithm~\ref{alg:SigGen} be $n_1=5$, $n_2=5$,
$m=5$, $m_0=1$, so that we have 1,024 signatures of length $2^m=32$.
Suppose for simplicity node $0$ has only two neighbors, whose
on-off signatures are $\B{S}_1$ and $\B{S}_2$, respectively:
\begin{align}
 \label{eq:4}
\B{S}_1 &= [0, j, 0, 0, 1, 0, 0, 0, -1, -j, 0, 1, 1, 0, 0, 0, \nonumber \\
& 0, -j, 0, 0, -1, 0, 0, 0, 1, j, 0, 1, -1, 0, 0, 0]\\
\B{S}_2 &= [1, j, 0, -j, 0, 0, 1, 0, 1, 0, -1, 0, 0, j, -1, -j, \nonumber \\
& {-1}, -j, 0, -j, 0, 0, 1, 0, 1, 0, 1, 0, 0, j, 1, j]
\end{align}
where the zeros in the signatures are due to erasures.
Suppose the channel gains are $U_1=3$ and $U_2=2j$.
In absence of noise, node 0 observes the signal
$U_1\B{S}_1+U_2\B{S}_2$ through its own off-slots as:
\begin{align}
&\tilde{\B{Y}} = %U_1 \B{S}_1 + U_2 \B{S}_2 =
[2j, 0, 0, 2, 0, 0, 0, 0, 0, 0, -2j, 3, 0, -2, -2j, 2, \nonumber \\
& \;\; {-2j}, 0, 0, 2, 0, 0, 0, 0, 0, 0, 2j, 3, 0, -2, 2j, -2].
\label{eq:Yt}
\end{align}}%
\begin{comment}
where the zeros correspond to node 0's own transmissions.
 Let the parameters of signature generation in
Algorithm~\ref{alg:SigGen} be $n_1=5, n_2=5, m=5, m_0=1$, so that we
have $n=10$ .
The on-off signatures of the two neighbors and the received signal $\tilde{\Y}$ with erasures at node $0$ are given below. Note the signatures are generated by RM codes with 50\% random erasures as in Algorithm~\ref{alg:SigGen}.]
\end{comment}
% In light of the original chirp algorithm, we emphasize the key steps
% as follows. For better illustration, we also provide a specific
% example in braces to show the results of intermediate steps.
%\DG{Given $\tilde{\B{Y}}$}, the key steps of the decoding procedure is described as follows.
\LZ{Given $\Y_r$ is initialized to $\tilde{\Y}$, the key steps of Algorithm~\ref{alg:Chirp} leading to the discovery of the first neighbor is described as follows:}
\begin{enumerate}
%\item Initialize the residual signal $\Y_r$ to $\tilde{\Y}$;
%[As an example just for illustration, let the parameters of signature generation in Algorithm~\ref{alg:SigGen} be $n_1=5, n_2=5, m=5, m_0=1$. Consider a special deployment of $2^{10}=1,024$ nodes in the network
%such that node $0$ has only two neighbors. The on-off signatures of the two neighbors and the received signal $\tilde{\Y}$ with erasures at node $0$ are given below. Note the signatures are generated by RM codes with 50\% random erasures as in Algorithm~\ref{alg:SigGen}.]
\item \LZ{Steps 7 to 12:} \\
  Note that $m_0=1$ in this case.  Take the Hadamard transform of the
  auto-correlation function of
  $\Y_r$ and its shift by $2^{m-1}$ to expose the chirps
 in the frequency domain, so that the first row of
  $\B{P}(\B{c})$ can be recovered, and then the entire matrix
  can be determined.
  \DG{Using $\tilde{\B{Y}}$ given by~\eqref{eq:Yt}, the index of the
    highest peak is the 21st. Therefore, the first row of $\B{P}(\B{c})$ is
    the binary representation of 20, i.e., the binary string of
    10100.  The matrix $\B{P}(\B{c})$ can then be uniquely determined.
    }

  \item \LZ{Steps 13 and 14:} \\
  Compute $\S^0(\B{a})=\expb{j\pi\left(\frac12
      \B{a}\tran{}\B{P}(\B{c})\B{a} \right)}$ for all $\B{a} \in
  \mathbb{Z}_2^m$ and apply Hadamard transform to the pointwise
  product of $\Y_r$ and the conjugate of $\S^0$ to recover $\B{b}$.
  \DG{In the example, the index of the
  highest peak is the 19-th in the first iteration, hence $\B{b}=\,$10010. }

\item \LZ{Steps 15 to 18:} \\
  Recover the erased signature $\S$ by pointwise product of
  $\phi_{\B{b},\B{c}}$ and $\B{r}$, then put together all
  signatures already recovered to form a matrix $\tilde{\SSS}$, \LZ{where all rows of $\tilde{\SSS}$ corresponding to the on-slots of node 0 are set to zero.} Determine the value of $\X$ which minimizes $\|\Y_r-\tilde{\SSS}\X\|_2$.
  %, where
  %$\tilde{\S}$ is constructed by setting all rows of $\S$ but those
  %which correspond to the off-slots of node 0 to zero, and then update
  %all corresponding coefficients.
  \DG{In the example, the reconstructed signature in the first
    iteration corresponds to the signature of the first neighbor
    ($\S_1$) and the corresponding coefficient $X_1$ is estimated to
    be $3.17+0.17j$, which is close to $U_1$.}

\begin{comment}
\item Update the residual signal and repeat steps 1 to 4 \DG{to
    discovery more nodes,} until the
  residual falls below a threshold or the total number of iterations
  \DG{reaches the maximum number of iterations as one desire.}
  %exceeds a threshold.
  %In the example the algorithm will try to identify $\S_2$.
  \end{comment}
\end{enumerate}
\LZ{The preceding steps are repeated to discover more nodes.  The
  algorithm terminates if either the total number of iterations
  reaches the maximum number of iterations as one desire, or among the
  discovered nodes, enough of them correspond to very weak
  coefficients, which implies that the algorithm starts to produce non-neighbors.}

%Algorithm~\ref{alg:Chirp} provides enough details about the scheme, which allows someone without familiarity with RM code to implement it.

%In the chirp reconstruction algorithm described
%in~\cite{HowCal08CISS}
\DG{We now justify the special scheme for generating the erasures in Algorithm 3.}
In order to recover the $i$-th row of the $m\times m$ symmetric
matrix corresponding to the largest energy component in the
residual signal, the auto-correlation is computed between the residual
signal of length $2^m$ and its shift by $2^{m-i}$. It is advisable to
guarantee that the positions of erasures in the received signal and
its shift are perfectly aligned as designed in Algorithm~\ref{alg:SigGen}.

\subsection{A Numerical Example}
\label{s:example}

We illustrate the performance of discovery using RM codes through
the following example.  The same network model is assumed as in
Section~\ref{s:optimal}, where there are $2^{20}$ valid NIAs.

\begin{figure}[t]
 \centering
 \includegraphics[width=5in]{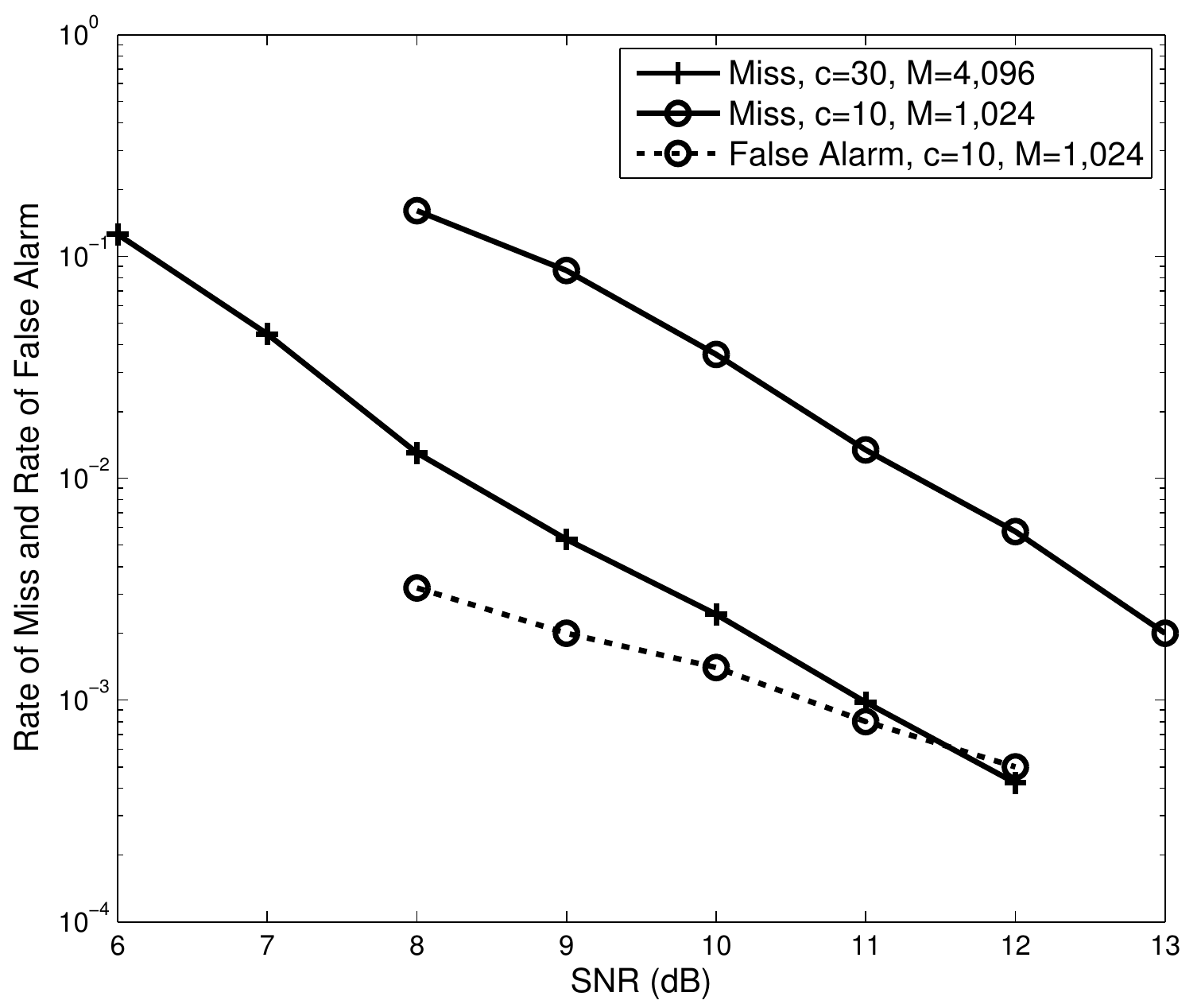}
 \caption{The rates of miss and the rate of false alarm versus SNR.}
\label{f:rm}
\end{figure}

First let the density of the nodes be such that each node has on
average $c=10$ neighbors.  Choose $m=n_1=n_2=10$, then the signature
length is $2^m=$ 1,024.  Averaged over 10 network realizations of the
large network, the rate of miss and the rate of false alarm of a
  total 1$0\times100=1$,000 nodes (with approximately 10,000
  neighbors in total) are plotted in Fig.~\ref{f:rm} against the
SNR.  Note that there are no false alarms registered during the
simulation when SNR is larger than \ZH{12}~dB.  We find that the total
error rate can be lower than 0.2\% at 13~dB SNR. In contrast, if
random on-off 1,024-bit signatures are used instead (see
Fig.~\ref{f:Pe10}), at least 26~dB SNR is needed to achieve the same
error rate, even if the size of the address space is only 10,000.

We repeat the simulation with the number of
average neighbors changed to $c=30$ and the parameters changed to
$m=n_2=12$, and $n_1=8$. In this case, the signature length is
$2^m=4$,096. During all 10 network realizations, there are no false
alarms and the total error rate can be lower than 0.2\% at 11~dB SNR.

\begin{figure}[t]
 \centering
 \includegraphics[width=5in]{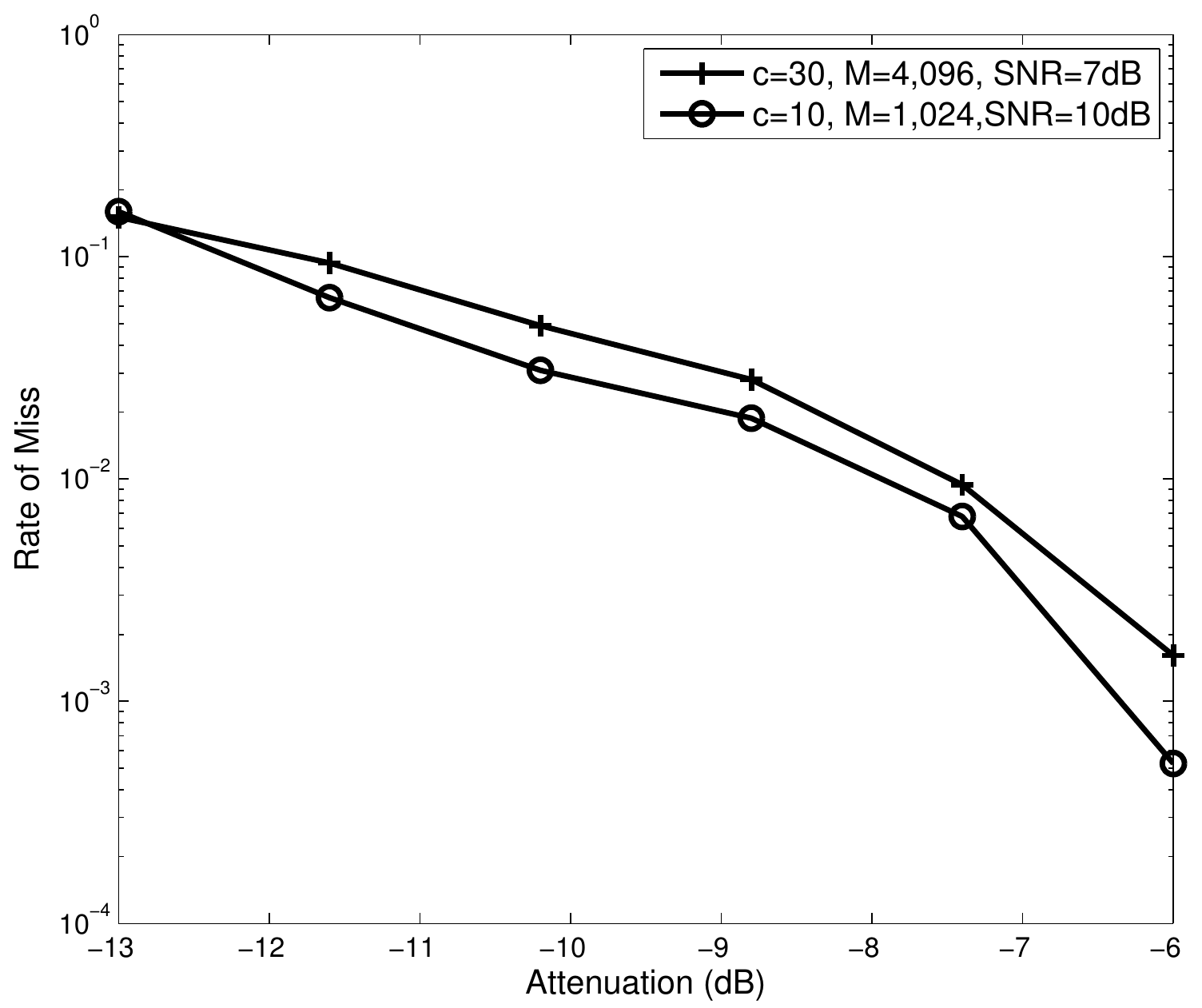}
 \caption{The rate of miss versus attenuation.}
\label{f:mr}
\end{figure}

\ZH{In order to show that the chirp decoding algorithm is highly resilient to the near-far problem, we demonstrate in Fig.~\ref{f:mr} that strong neighbors will be detected with very high probability so that their interference to weaker neighbors can be removed. In the case of average $c=10$ neighbors, when the signature length is 1,024 and SNR is 10~dB, we can see that the rate of miss decreases as the neighbors become stronger, and the rate of miss is below 0.1\% at $-6$~dB attenuation. The simulation is repeated with the number of average neighbors changed to $c=30$, the length of signature changed to 4,096 and SNR changed to 7~dB. We can see that all neighbors with attenuation less than $-6$~dB are successfully discovered with miss rate less than 0.2\%.}

\begin{comment}
For comparison, if random on-off 2,048-bit signatures are used instead
(see Fig.~\ref{f:Pe30}), at least 29~dB SNR is needed to achieve the
same error rate. Although the signatures constructed from RM codes are
longer, the RM-based scheme is more desirable in practice because the
size of the address space is much larger and the required SNR is much
lower.
\end{comment}

\section{Comparison with Random Access}\label{s:Compare}

We compare the performance of the compressed neighbor discovery
schemes described in Sections~\ref{s:onoff} and~\ref{s:rm} with that
of conventional random-access discovery schemes.
Only one frame interval is needed by compressed neighbor discovery, as
opposed to many frames (often in the hundreds) in case of random
access.  Thus compressed neighbor discovery also offers significant
reduction of synchronization and error-control overhead embedded in every frame.

\subsection{Comparison with Generic Random-Access Discovery}
\label{sec:CompareBLT}

Suppose a random-access discovery scheme is used, such as the
``birthday'' algorithm in \cite{McGBor01MANC}.  Nodes contend to
announce their NIAs over a sequence of $k$ contention periods.  In
each period, each neighbor independently chooses to either transmit
(with probability $\theta$) or listen (with probability
$1-\theta$).   Let $\rho=c/\len$.   The error rate is equal to the
probability of one given neighbor being missed, which is given by
\begin{equation} \label{eq:aloha}
%  \frac{1}{c}
  \sum_{z=1}^{\len}{\binom{\len}{z}} \rho^z
  \left(1- \rho \right)^{\len-z}
  %\left( \frac{c}{\len} \right)^z
    % \left(1- \frac{c}{\len} \right)^{\len-z}
    \left[ 1 - \theta \left(1- \theta \right)^{z-1}\right]^k \ .
\end{equation}
%where $\rho=c/\len$. % and there are $\ctp$ contention periods.

Consider a network with 10,000 nodes, so in each contention
  period, the number of bits transmitted is at least $\lceil
  \log_2(10^4) \rceil = 1$4 just to carry the NIA. For fair comparison with
  compressed neighbor discovery schemes, we assume time is slotted and
  QPSK modulation is used.
Table~\ref{t:compare_GT} lists the amount of transmissions needed by random
access discovery according to~\eqref{eq:aloha} and by compressed
discovery based on 2-tolerance group testing
%Algorithm~\ref{alg:Multiple} with 2-tolerance test
(see Figs.~\ref{f:Pe30} and~\ref{f:Pe10}) in order to achieve the
target error rate of 0.002 in cases of 10 and 30 neighbors.

Evidently, random-access discovery requires hundreds of 14-bit frame
transmissions to guarantee the same performance achieved by compressed
discovery using a single frame transmission.  The latter scheme uses
much longer frames.  Still, the total number of symbols required by
compressed discovery is substantially smaller, and in fact the
advantage is greater in case of more neighbors.
% We find that group testing based neighbor discovery significantly
% outperforms random access scheme, especially in case of a relatively
% large number of neighbors.

\begin{table}[t]
\begin{center}
\setlength{\tabcolsep}{18pt}
\caption{Comparison between random-access discovery and compressed
  neighbor discovery based on group testing.}
\begin{tabular}{|c|c|c|}
\hline
       & random access & group testing \\
\hline
$c=10$
& 194 frames & 1 frame \\
& 1,358 symbols & 1,024 symbols \\
\hline
$c=30$
& 534 frames & 1 frame \\
& 3,738 symbols & 2,048 symbols \\
\hline
\end{tabular}
\label{t:compare_GT}
\end{center}
\end{table}

Similar comparison can be made between random-access discovery and
compressed discovery based on RM codes.  Consider a network with
$2^{20}$ nodes.  To achieve the target rate of 0.002 in case of 10 or
30 neighbors, Table~\ref{t:compare_RM} lists the amount of
transmissions needed by random-access discovery according
to~\eqref{eq:aloha} and by compressed discovery based on RM codes with
chirp decoding (see Fig.~\ref{f:rm}).
Again, compressed discovery significantly outperforms random-access
discovery.

\begin{table}[t]
\begin{center}
\setlength{\tabcolsep}{18pt}
\caption{Comparison between random-access discovery and compressed
  discovery based on RM codes.}
\begin{tabular}{|c|c|c|}
\hline
       & random access & RM codes \\
\hline
$c=10$
& 194 frames & 1 frame \\
& 1,940 symbols & 1,024 symbols \\
\hline
$c=30$
& 534 frames & 1 frame \\
& 5,340 symbols & 4,096 symbols \\
\hline
\end{tabular}
\label{t:compare_RM}
\end{center}
\end{table}

The efficiency of compressed neighbor discovery can be
significantly higher than that of random access if all overhead is
accounted for.   This is because that sending a 14-bit or 20-bit
NIA reliably over a fading channel may require up to a hundred symbol
transmissions or more.  We believe using compressed discovery can
reduce the amount of total discovery overhead by an order of
magnitude.

\subsection{Comparison with IEEE 802.11g}

It is also instructive to compare compressed neighbor discovery with
the popular IEEE 802.11g technology.  Consider the ad hoc mode of
802.11g with active scan, which is basically a random-access discovery
scheme.  The signaling rate is 4~$\mu s$ per orthogonal frequency
division multiplexing (OFDM) symbol.  One
%contention period has to cover the transmission time for one
% (* 850 194)
probe response frame takes about 850~$\mu s$.  (The response
  frame includes additional bits but is dominated by the NIA.)  Thus
it takes at least $850$ $\mu s \times 194 \approx 165$ $ms$ for a
query node to discovery $10$ neighbors with error rate $0.002$ or
lower.  If compressed neighbor discovery with on-off signature
is used, 1,024 symbol
transmissions suffice to achieve the same error rate.  Using
802.11g symbol interval (4~$\mu s$), reliable discovery takes merely
4.1~$ms$.  A highly conservative choice of the symbol interval is $30$
$\mu s$, which includes carrier (on-off) ramp period (say $10$ $\mu
s$) and the propagation time (less than 1 microsecond for 802.11
range).  Compressed neighbor discovery then takes a total of
%  $1,000 \times 30$ $\mu s \approx
$30$ $ms$, less than $1/5$ of that required by 802.11g.

% (/ 30.0 96.05)

\section{Concluding Remarks}\label{s:con}

In this paper, we have developed two compressed neighbor discovery
schemes, which are efficient, scalable, and easy to implement. The
on-off signaling used in neighbor discovery schemes was first proposed
in~\cite{GuoZha10Allerton} and referred to as {\em rapid
  on-off-division duplex}, or {\em RODD}. Such signaling departs from the
collision model and fully exploits the superposition nature of the
wireless medium~\cite{ZhaGuo11ISIT}. Moreover, using on-off signatures
allows half-duplex nodes to achieve network-wide full-duplex
discovery.  It is interesting to note that transmission of pulses by
each node (which identifies the node) is scheduled at the symbol
level, rather than at the timescale of the frame level.

\LZ{The neighbor discovery problem is different from most other
  applications of compressed sensing in the literature because of the
  sheer scale of the problem. The number of unknowns is typically
  $2^{20}$ or more. We choose to use RM codes because of its
  scalability and effectiveness for compressed sensing. At this point,
  there are no other practical codes which are known to deliver
  comparable performance for noisy compressed sensing at this scale
  and efficiency.}

A brief discussion of how neighbor discovery is triggered is in
order.  If a single node (e.g., a new comer) is interested in its
neighborhood, it may send a query message, so that only the neighbors
which can hear the message will respond immediately.
To implement network-wide discovery, nodes can be programmed to
simultaneously transmit their on-off signatures at regular,
pre-determined epochs, so that all nodes discover their
respective neighbors.  This also prevents neighbor discovery
from interfering with data transmission.

Compressed neighbor discovery is well suited and in fact significantly
outperforms existing schemes for mobile networks where the topology of
the network changes over time.
\DG{Depending on the mobility, it may be desirable to carry out neighbor
  discovery periodically.  If this is done frequently, the toplogy may
  not change much, hence}
%(Note that within a single frame interval ($<100$~$ms$) nodes move by
%a small amount.)
it is also possible to create a Markov model for connectivity and
incorporate the model into the neighborhood inference problem.
This is left to future work.

Finally, we note that very recently Qualcomm has developed the FlashLinQ
technology based on OFDM,
which carries out neighbor discovery over a large number of orthogonal
time-frequency slots~\cite{NiSri10SigMet}.  Over each slot, however,
the scheme is still based on random access.  The
schemes proposed in this paper can also be extended to multicarrier systems.
This is also a direction for future work.

% Nevertheless, network-wide synchronization is required before
% network-level neighbor discovery takes place, which is not hard to
% achieve by utilizing network consensus algorithms~\cite{CaoMor08SIAM}.
% %Note that in~\cite{VasTow09TechUMass}

% MAY ADDRESS THE ISSUE THAT NETWORK CONSENSUS IS REQUIRED IN BOTH
% THIS WORK AND Section 7.1 IN~\cite{VasTow09TechUMass}.

\section*{Acknowledgement}

We thank Robert Calderbank and Sina Jafarpour
for %useful
 discussion and for sharing codes of the chirp decoding
algorithm.
We also thank Kai Shen for assistance in carrying out some simulations
in Section~\ref{s:onoff}.
%  We also thank the anonymous reviewers for their constructive comments.

\bibliographystyle{IEEEtran}
\bibliography{def,dguo,alljab,neighbor_110826}
\end{document}